\documentclass[1p,12pt]{elsarticle}
\usepackage{amssymb}
\journal{Nuclear Physics B}
\begin{document}
\begin{frontmatter}
\title{Quantum phase diagram of the half filled Hubbard model with bond-charge interaction.
}

\author[AA]{A. O. Dobry \corref{PPP}}, \ead{dobry@ifir-conicet.gov.ar}
\author[BB]{A. A. Aligia }, 

\address[AA]{Facultad de Ciencias Exactas Ingenier\'{\i}a y Agrimensura, Universidad Nacional de Rosario
and Instituto de F\'{\i}sica Rosario, Bv. 27 de
Febrero 210 bis, 2000 Rosario, Argentina}
\address[BB]{Centro At\'{o}mico Bariloche and Instituto Balseiro, Comisi\'{o}n Nacional
de Energ\'{\i }a At\'{o}mica, 8400 Bariloche, Argentina}

\cortext[PPP]{Corresponding author. Fax: 54 341 4821771}

\begin{abstract}
Using quantum field theory and bosonization, we determine the quantum phase diagram 
of the one-dimensional Hubbard model with bond-charge interaction $X$ in addition to the usual Coulomb
repulsion $U$ at half-filling, for small values of the interactions. 
We show that it is essential to take into account formally irrelevant terms of order $X$.
They generate relevant terms proportional to $X^2$ in the flow of the renormalization
group (RG). These terms are calculated using operator product expansions.
The model shows three phases separated by a charge transition at $U=U_c$ and a spin transition
at $U=U_s>U_c$. For $U<U_c$ singlet superconducting correlations dominate, while for $U>U_s$, 
the system is in the spin-density wave phase as in the usual Hubbard
model. For intermediate values $U_c<U<U_s$, the system is in a spontaneously
dimerized bond-ordered wave phase, which is absent in the ordinary Hubbard model with $X=0$.
We obtain that the charge transition remains at $U_c=0$ for $X \neq 0$. Solving the RG equations
for the spin sector, we provide an analytical expression for $U_s(X)$. The results, with
only one adjustable parameter, are in excellent 
agreement with numerical ones for $X < t/2$ where $t$ is the hopping.
\end{abstract}

\begin{keyword}
Bond-charge interaction \sep Correlated hopping \sep Renormalization group \sep Hubbard model  \sep Operator product expansion  
\PACS 71.10.Fd \sep 71.10.Hf \sep 71.10Pm \sep 11.10.Hi 
\end{keyword}

\end{frontmatter}

\section{Introduction}

\subsection{The model}

\label{1p1}

The Hubbard model, with nearest-neighbor hopping $t$ and on-site repulsion $U
$ has been widely used to study the effects of correlations, as a simplified
model to describe compounds of transition metals and other systems. However,
one expects that in any system, in general the hopping between two sites
depend on the occupation of these two sites, which leads to the presence of
bond-charge interactions (also called correlated hopping terms) in the
Hamiltonian [such as $X$ in Eq. (\ref{hamil}) and $\sum_{\sigma ,\langle
ij\rangle }(c_{i\sigma }^{\dagger }c_{j\sigma }+{\rm H.c.})n_{i-\sigma
}n_{j-\sigma }$]. For example, in simple systems with one relevant orbital
per site, one would expect that when electrons are added to one site, the
screening of the core charge increases, and as a consequence, the wave
function of the orbital expands and the hopping to the nearest sites
increases. In general, any one-band effective model derived from more
complex Hamiltonians to describe the low-energy physics of some system, contains bond-charge interactions. 
In fact, the generalized Hubbard model with correlated hopping terms has been derived and used to
describe the low-energy physics of intermediate valence systems \cite{foglio}, 
organic compounds \cite{kive,baeri,gamm,zhang,stra,br}, a Hubbard model
including lattice vibrations \cite{phon}, cuprate superconductors \cite%
{schu,simon,opt}, and more recently optical lattices \cite{duan,good,kest}.
This is particularly interesting because the parameters can be tuned
experimentally in a wide range \cite{kest,duan2}. First-principles
calculations in transition-metal complexes suggest that the correlated
hopping terms can be large \cite{impu}.

As we shall see, the presence of bond-charge interaction leads to
qualitatively new physics. One example is that in two dimensions $d$-wave
pairing correlations, which are already present in the Hubbard model \cite%
{scala} are strongly enhanced in the generalized Hubbard model for the
cuprates \cite{lili}, and one obtains $d$-wave superconductivity already at
the mean-field level \cite{dw}. In one dimension (1D), field-theoretical 
\cite{jaka,bos} and numerical \cite{bos,topo} results show the presence of a
spontaneously dimerized bond-ordering wave (BOW) and a phase with dominant
triplet superconducting correlations at large distances that are absent in
the ordinary Hubbard model. For special values of the parameters, the model
with two- and three-body interactions has been solved exactly by the Bethe
ansatz \cite{igor}.

The simplest model with bond-charge interaction has been proposed by Hirsch
motivated by his theory of hole superconductivity \cite{hirsch,hir2}. The
Hamiltonian can be written as 
\begin{equation}
H=-t\sum_{\sigma =\uparrow ,\downarrow ,\langle ij\rangle }(c_{i\sigma
}^{\dagger }c_{j\sigma }+{\rm H.c.})+U\sum_{i}n_{i\uparrow }n_{i\downarrow
}+X\sum_{\sigma ,\langle ij\rangle }(c_{i\sigma }^{\dagger }c_{j\sigma }+%
{\rm H.c.})(n_{i-\sigma }+n_{j-\sigma }).  \label{hamil}
\end{equation}%
In 1D, the model can display a phase with dominant singlet superconducting
(SS) correlations, even for positive $U$ \cite{bos,japa,tll,bulka}. For $X=t$%
, the model has been solved exactly and there is a metal-insulator
transition for increasing $U$ \cite{exac,boer,flux}. More recently, the role of
entanglement in this quantum phase transition has been studied \cite{entro,agm}. 
The response of the system to an applied magnetic flux
indicates that the metallic phase is not superconducting \cite{flux}.
However, the ground state is highly degenerate in this phase at it is
difficult to predict from the exact solution what happens when the degeneracy is
lifted by a small but finite $X-t$. In any case, standard bosonization \cite%
{jaka,bos,japa} and numerical studies \cite{tll} have provided a general
physical picture of the behavior of the model, except at half filling. In
this case, taking as usual only the leading terms in the lattice constant $a$%
, $X$ disappears in the bosonization treatment [it enters as $X\cos (\nu \pi
)$, where $\nu $ is the filling fraction \cite{japa,bos}]. Therefore,
standard field theory predicts a SS phase for $U<0$, and a spin-density wave
(SDW) phase for $U>0$, as in the usual Hubbard model. However, a charge
insulator-metal transition driven by $X$ at finite $U_{c}>0$ has been found
numerically \cite{sup-ins} and later a quantum phase diagram has been
derived which includes a BOW phase \cite{cola}. 
Recently, new numerical studies of the model at arbitrary filling identify 
regions of phase separation for $X > 0.5 t$ \cite{ari2}.

\subsection{Phase diagram at half filling}

\label{1p2}

\begin{figure}[tbp]
\includegraphics[width=14cm]{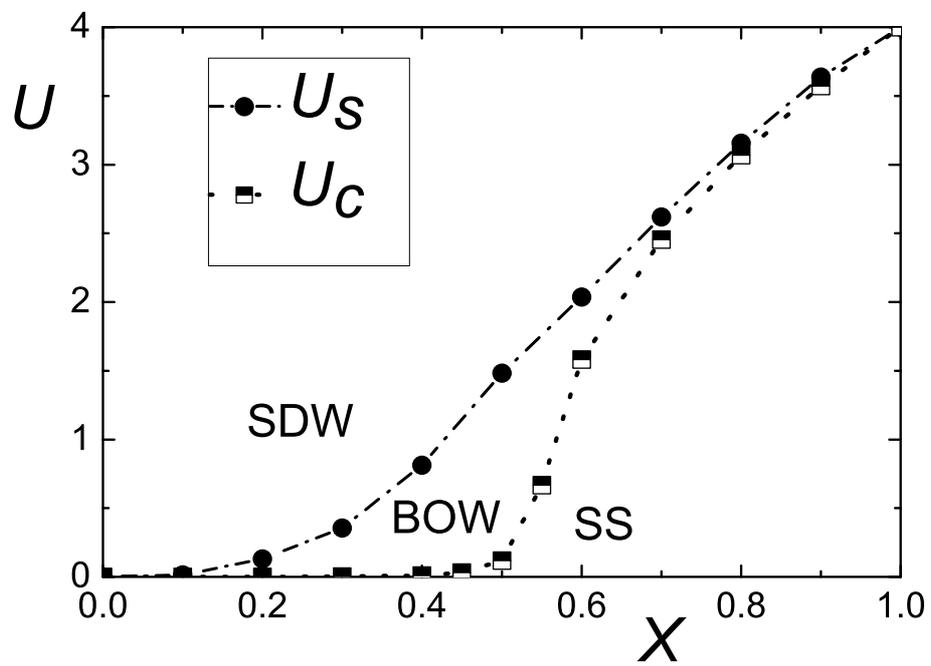}
\caption{Phase diagram of the model at half filling obtained from the method
of level crossings. Squares: charge transition. Solid circles: spin
transition. The unit of energy is taken as $t=1$.}
\label{pd}
\end{figure}

The quantum phase diagram for $\nu =1/2$ has been obtained by a combination
of different numerical techniques \cite{cola,ari2}. In Fig. \ref{pd} we reproduce
the phase diagram obtained by the method of topological transitions 
\cite{topo,abb}. These transitions correspond to jumps in the charge and spin
Berry phases which signal the corresponding transitions between the
thermodynamic phases, and coincide with a corresponding crossing of excited
levels, justified on the basis of conformal-field theory \cite{abb,naka}.
These quantities are also related to charge and spin localization indicators 
\cite{loc1,loc2,pss,toro} used for example to characterize valence-bond-solid
states in quantum spin chains \cite{nato}. While changes in the Berry phase
are proportional to changes in polarization, the spin Berry phase tensor
provides a geometric characterization of the ferrotoroidic moment \cite{toro}.

The critical values of $U$ for the charge ($U_{c}$) and spin ($U_{s}$)
transition have been calculated in systems of up to $L=14$ lattice sites,
and extrapolated to the thermodynamic limit using a parabola in $1/L^{2}$.
Fig. \ref{pd} displays the extrapolated values for $0\leqslant X\leqslant 1$. 
It is important to note that this interval can be extended to the whole
real axis using symmetry properties of the Hamiltonian \cite{tll}. A change
of phase of half of the sites [$c_{i\sigma }^{\dagger }\rightarrow
(-1)^{i}c_{i\sigma }^{\dagger }$], interchanges the signs of $t$ and $X$, so
that $H(-t,-X,U)\equiv H(t,X,U)$. Combining this with an electron-hole
transformation ($c_{i\sigma }^{\dagger }\rightarrow c_{i\sigma }$), one
obtains at half filling

\begin{equation}
H(t-2X,-X,U)\equiv H(t,X,U)\equiv H(2X-t,X,U).  \label{sym}
\end{equation}%
Then, if the critical ratio $u(x)$ ($U_{c}/t$ or $U_{s}/t$) with $x=X/t$, $t>0$, 
is known in the interval $0\leqslant x<1/2$, it can be
extended to negative values of $X$ using the first Eq. (\ref{sym}).
Similarly, the second Eq. (\ref{sym}) maps the interval $1/2<x\leqslant 1$
onto $x\eqslantgtr 1$. Explicitly

\begin{equation}
{\rm if }x<0{\rm , \; }u(x)=(1-2x)u\left( \frac{-x}{1-2x}\right) {\rm ; \; if }%
x\eqslantgtr 1{\rm , \; }u(x)=(2x-1)u\left( \frac{x}{2x-1}\right).
\label{rel}
\end{equation}%
It is interesting to note that the end points $x\rightarrow \pm \infty $ are
mapped onto $x=1/2.$

For $U>U_{c}$ ($U<U_{s}$), the system has a charge (spin) gap. For $U<U_{c}$%
, the system is in the SS phase, while for $U>U_{s}$, the system is in the
SDW phase, according to the dominant correlation functions at large
distances. In between, for $U_{c}<U<U_{s}$, one has the fully gapped BOW
phase. For small values of the interactions, the dominant correlation
functions in each phase can be understood from field theory \cite{jaka,bos}.
For $X>t/2$, the SS phase displays incommensurate correlations \cite{cola},
which can be qualitatively understood using a mean field approximation in
one of the terms obtained from bosonization \cite{cola}, leading to a
commensurate-incommensurate transition, with some similarities to the
physics of the Hubbard model when a large next-nearest-neighbor hopping is
added \cite{ttpu}, and some spin systems \cite{nerse}. The spontaneously
dimerized BOW phase has also been also found in the Hubbard model with
alternative on-site energies \cite%
{abb,fab1,fab2,torio,manma,rap,bat1,dimer,tinca}, where it displays
ferroelectricity \cite{abb,bat1,dimer}. 
The presence of the BOW phase in this model was 
first predicted using field theory
and bosonization \cite{fab1,fab2}.

For $X<t/2$, the numerical results for the charge transition are consistent
with $U_{c}=0$, as in the ordinary Hubbard model \cite{cola}. The accuracy
of the results are not enough to establish if there is a kink or not at $%
X=t/2$, $U_{c}=0$. Arianna Montorsi has found that a good fit of the
numerical results for $t/2\leqslant X\leqslant t$ is \cite{ari}

\begin{equation}
{\rm if }x\eqslantgtr 1/2{\rm , \; }u_{c}=4\sqrt{2x-1}{\rm , \; }  \label{am}
\end{equation}%
which is consistent with a kink at $x-1/2=u_{c}=0$, and has the nice
property that when it is extended analytically to $x\eqslantgtr 1$, it
satisfies the second symmetry relation (\ref{rel}). The value $U_{c}=4t$ for 
$X=1$, is consistent with the exact solution \cite{exac,boer,flux} To our
knowledge, no justification of Eq. (\ref{am}) exists so far.

It has been verified that the spin transition is of Kosterlitz-Thouless
type \cite{cola}. In contrast to $U_{c}$, the critical value for the spin
transition $U_{s}(X)$ represented in Fig. \ref{pd} is smooth. For small
values of $X$, it increases as $X^{2}$, while for $X\sim t/2$
there is an inflection point. For $X<t$, $U_{s}>U_{c}$. While for $X=t$, $%
U_{s}=U_{c}=4t$, the second Eq. (\ref{rel}) implies that for $X>t$, also $%
U_{s}>U_{c}$. Therefore, there is no crossing between $U_{c}(X)$ and $%
U_{s}(X)$ at $X=t$. The fact that there is a finite value of $U_{s}(1/2)$
and the second Eq. (\ref{rel}) imply that $U_{s}(X)$ grows linearly with $X$
for $X\rightarrow +\infty $, in contrast to the $\sqrt{X}$ behavior for the
charge transition predicted by Eq. (\ref{am}).

\subsection{Previous field-theoretical results}

\label{1p3}

As stated in Section \ref{1p1}, standard continuum limit field theory and
bosonization fails at half filling because $X$ disappears from the $g_{i}$
coupling constants \cite{japa,bos}. In Ref. \cite{cola} we have calculated
vertex corrections to these $g_{i}$ using second order perturbation theory
in the bond-charge interaction $X$. The approach is similar to that done by
Tsuchiizu and Furusaki for the Hubbard model extended with nearest-neighbor
repulsion \cite{japan}, but for our Hamiltonian, Eq. (\ref{hamil}) it is not
necessary to introduce a low-energy cutoff. This approach led to the
following critical $U$ at the spin transition

\begin{equation}
U_{s}=\frac{8X^{2}}{\pi (t-X)}  \label{usve}
\end{equation}%
This function lies below the numerical points in Fig. \ref{pd}, but seems to
represent correctly the limit $X\rightarrow 0$. However, unfortunately the
prediction of this approach for the charge transition is $U_{c}\sim U_{s}/2$
for small $X$, instead of $U_{c}=0$ found numerically. In addition, with
vertex corrections only, it is not possible to explain the nature of the
incommensurate SS phase for $X>t/2$. 

  In Ref. \cite{cola} we have also considered in the bosonized theory,
a term in next to
leading order in the lattice parameter $a$, which couples charge and spin in
a mean-field approximation. However, this approximation is questionable, and
the quantitative agreement between the analytical and numerical results for the charge
transition is poor.

In this work we include all terms of next to leading order in $a$, and
include them in a renormalization group (RG) treatment. This approach is
superior to perturbation theory in $X$ (as included in Ref. \cite{cola}
through vertex corrections). Retaining all these terms leads to a lengthy
algebra, but unfortunately selecting only a few of them, breaks the SU(2)
symmetry and leads to wrong results. Since our approach is a weak coupling
one, we restrict our study to $X<t/2$, which seems to be the more realistic
regime of parameters. Our effort is rewarded by an excellent agreement with
the numerical results for both critical values of $U$ at the corresponding
transitions.

\section{The field-theoretical approach}

\subsection{The continuum limit}

\label{2p1}
In order to construct the low-energy field theory for the Hamiltonian Eq. (%
\ref{hamil}), we suposse that both $U$ and $X$ are small. Therefore,
in the Fourier development of fermion operators we retain only the modes near $-k_F$ and
$k_F$, 
where $k_F$ is the Fermi wave vector. Introducing a cutoff $\Lambda<< 1/a$, 
where $a$ is the lattice parameter, and calling 
$L$ the lenght of the system, the local annihilation operator $c_{n \sigma}$ can be written as:
\begin{eqnarray}
c_{n \sigma}&=&\sqrt{\frac{a}{L}} \sum^{\frac{\pi}{a}}_{k=-\frac{\pi}{a}} e^{ik n a} c_{k\sigma}\nonumber\\&\sim& \sqrt{\frac{a}{L}}\left[ e^{-ik_F na} \sum_{-\Lambda<k+k_F<\Lambda}  e^{i(k+k_F)n a} c_{k\sigma}+
e^{ik_F na} \sum_{-\Lambda<k-k_F<\Lambda}  e^{i(k-k_F)n a} c_{k\sigma}\right]\equiv
\nonumber\\
&&
\sqrt{a} \left[ e^{-ik_F n}
\psi_{\sigma-}(x=na)+e^{ik_F na} \psi_{\sigma+}(x=na)\right]  
\label{fermion}
\end{eqnarray}
in the last step we have introduced
the left and right fermionic fields $\psi_{\sigma-}(x)$ and $\psi_{\sigma+}(x)$ respectively, by replacing the discrete lattice index $n$ by a continuous variable $x \sim na$. 
This is possible because of the very small change undergone by sums in the second line of the previous equation,
when one goes from site $n$ to $n+1$. 

Now we can undertake a gradient expansion for $H$ by making the replacement
\begin{eqnarray}
 \psi_{\sigma\pm}[x=(n+1)a)]\rightarrow\psi_{\sigma\pm}(x=na)+ a
\partial_x \psi_{\sigma \pm}(x))
\label{discretder}
\end{eqnarray}
in all the terms of Eq. (\ref{hamil}). For the hopping operator we obtain
\begin{eqnarray}
(c_{n\sigma }^{\dagger }c_{n+1\sigma }+c_{n+1\sigma }^{\dagger }c_{n\sigma
}) &\sim& 2ia(-1)^{n}(\psi _{\sigma -}^{\dagger }\psi _{\sigma +}-\psi
_{\sigma +}^{\dagger }\psi _{\sigma -})+ \nonumber \\
&a^{2}&i[(\psi _{\sigma +}^{\dagger }\partial _{x}\psi _{\sigma +}-\partial
_{x}\psi _{\sigma +}^{\dagger }\psi _{\sigma +}-\psi _{\sigma -}^{\dagger
}\partial _{x}\psi _{\sigma -}+\partial _{x}\psi _{\sigma -}^{\dagger }\psi
_{\sigma -})+  \nonumber \\
&(-1)^{n}&\partial _{x}(\psi _{\sigma -}^{\dagger }\psi _{\sigma +}-\psi
_{\sigma +}^{\dagger }\psi _{\sigma -})]+O(a^{3}),  \label{conthopp}
\end{eqnarray}%
where we have used $k_F=\pi /(2a)$. The number operator becomes
\begin{eqnarray}
n_{n{\sigma }} &\sim &
a \bigg[\rho_{\sigma +}(x)+\rho_{\sigma -}(x))
\nonumber\\
&+&(-1)^n\left(\psi _{\sigma +}^{\dagger
}(x)\psi _{\sigma -}(x)+\psi _{\sigma -}^{\dagger }(x)\psi
_{\sigma +}(x)\right)\bigg]  
\end{eqnarray}
with $\rho _{\sigma +}=\psi _{\sigma +}^{\dagger }\psi _{\sigma +}$ and $%
\rho _{\sigma -}=\psi _{\sigma -}^{\dagger }\psi _{\sigma -}$.

By replacing $\sum_n$ by $\int \frac{dx}{a}$ and taking into account that 
the integration of terms with an oscillating $(-1)^n$ prefactor vanish, 
one obtains for the Hubbard  Hamiltonian [corresponding to the first 
two terms of Eq. (\ref{hamil})] the following form:
\begin{eqnarray}
H_U&=&
i v_F \int dx \left\{\psi _{\sigma +}^{\dagger }\partial _{x}\psi _{\sigma +}-\partial
_{x}\psi _{\sigma +}^{\dagger }\psi _{\sigma +}-\psi _{\sigma -}^{\dagger
}\partial _{x}\psi _{\sigma -}+\partial _{x}\psi _{\sigma -}^{\dagger }\psi
_{\sigma -}\right\}+\nonumber\\
&&\int dx \sum_\sigma\bigg\{ \frac{g_{4\perp}}{2} (\psi^{\dagger}_{%
\sigma+}\psi_{\sigma+}\psi^{\dagger}_{\overline{\sigma}+}\psi_{\overline{%
\sigma}+} +\psi^{\dagger}_{\sigma-}\psi_{\sigma-}\psi^{\dagger}_{\overline{%
\sigma}-}\psi_{\overline{\sigma}-}) +g_{2\perp}
\psi^{\dagger}_{\sigma+}\psi_{\sigma+}\psi^{\dagger}_{\overline{\sigma}%
-}\psi_{\overline{\sigma}-} +  \nonumber \\
&&g_{1\perp} \psi^{\dagger}_{\sigma+}\psi_{\sigma-}\psi^{\dagger}_{\overline{%
\sigma}-}\psi_{\overline{\sigma}+}+ \frac{g_{3\perp}}{2} (\psi^{\dagger}_{%
\sigma-}\psi_{\sigma+}\psi^{\dagger}_{\overline{\sigma}-}\psi_{\overline{%
\sigma}+}+h.c.)\bigg\}.  \label{hhubgeo}
\end{eqnarray}
The first line corresponds to the usual free Dirac Hamiltonian with 
$v_F=at$ the bare Fermi velocity 
(later we shall use $v_F=a(t-X)$, the Hartree-Fock value \cite{bos}) 
The terms with prefactors $g_{1\perp }$, $g_{2\perp
}$, $g_{3\perp }$, $g_{4\perp }$ correspond to
backward, forward two branch, Umklapp and forward one branch, respectively.
While all these constants are equal to $a U$, they might run independently under a
renormalization group (RG) flow. Moreover, in units 
in which $\hbar=1$ and $v_F=1$, 
the couplings $g_i$ 
are dimensionless. Now, if a coupling constant $g_i$ has units $E^{d-\Delta_i}$, $\Delta_i$ is known as the scaling dimension of the corresponding operator $O_i$, where $d$
is the spacetime dimension ($2$ in our case) \cite{Polchinski}. Therefore all interactions in Eq. (\ref{hhubgeo})
have scaling dimension $2$, they are marginal operators. 
It is known that depending on the sign of the $g_i$ 
which correspond to the charge or spin sector of the theory, they can become
marginally relevant or irrelevant. The first case leads to a charge 
or spin gap \cite{giama} (see also Section \ref{2p4}).  
Let us see how this situation is modified by inclusion of the correlated hopping term 
[the last one in Eq. (\ref{hamil})]

The sum of the number operators in (\ref{hamil}) has the following gradient expansion:
\begin{eqnarray}
(n_{n\overline{\sigma}}+n_{i+1\overline{\sigma}})&\sim& 2 a (\rho_{\overline{%
\sigma} +}+\rho_{\overline{\sigma} -})+a^2 [\partial_x(\rho_{\overline{\sigma%
} +}+\rho_{\overline{\sigma} -})-\nonumber
\\ &(-1)^n& (\partial_x(\psi^{\dagger}_{%
\overline{\sigma}-}\psi_{\overline{\sigma}+}+ \psi^{\dagger}_{\overline{%
\sigma}+}\psi_{\overline{\sigma}-}))]+O(a^3)
\label{nini+1cont}
\end{eqnarray}
multiplying (\ref{conthopp}) by (\ref{nini+1cont}) we see that the terms quadratic in $a$ are oscillating
and vanish under integration. This is the result anticipated in Section \ref{1p1}.
This means that no scattering $\it at$ the Fermi level is generated by the correlated hopping interaction.
We should include term up to $O(a^2)$ in the Hamiltonian. We obtain:
 
\begin{eqnarray}
H_{X} &=&i\int dx\sum_{\sigma }\bigg\{g_{4\perp }^{\prime }[-\partial
_{x}\psi _{\sigma +}^{\dagger }\psi _{\sigma +}\psi _{\overline{\sigma }%
+}^{\dagger }\psi _{\overline{\sigma }+}+\partial _{x}\psi _{\sigma
-}^{\dagger }\psi _{\sigma -}\psi _{\overline{\sigma }-}^{\dagger }\psi _{%
\overline{\sigma }-}-{\rm H.c.}]  \nonumber \\
&&g_{2\perp }^{\prime }[-\partial _{x}\psi _{\sigma +}^{\dagger }\psi
_{\sigma +}\psi _{\overline{\sigma }-}^{\dagger }\psi _{\overline{\sigma }%
-}+\partial _{x}\psi _{\sigma -}^{\dagger }\psi _{\sigma -}\psi _{\overline{%
\sigma }+}^{\dagger }\psi _{\overline{\sigma }+}-{\rm H.c.}]  \nonumber \\
&&g_{1\perp }^{\prime }[-\partial _{x}\psi _{\sigma +}^{\dagger }\psi
_{\sigma -}\psi _{\overline{\sigma }-}^{\dagger }\psi _{\overline{\sigma }%
+}+\partial _{x}\psi _{\sigma -}^{\dagger }\psi _{\sigma +}\psi _{\overline{%
\sigma }+}^{\dagger }\psi _{\overline{\sigma }-}-{\rm H.c.}]  \nonumber \\
&&g_{3\perp }^{\prime }[-\partial _{x}\psi _{\sigma -}^{\dagger }\psi
_{\sigma +}\psi _{\overline{\sigma }-}^{\dagger }\psi _{\overline{\sigma }%
+}+\partial _{x}\psi _{\sigma +}^{\dagger }\psi _{\sigma -}\psi _{\overline{%
\sigma }+}^{\dagger }\psi _{\overline{\sigma }-}-{\rm H.c.}]\bigg\}
\label{hXgeo}
\end{eqnarray}%
where all $g_{i}^{\prime }=a^{2}X$.  The essential differences between the field
theory given by Eq. (\ref{hXgeo}) and the one given by Eq. (\ref{hhubgeo})
is the non local nature of the interaction arising
from the derivatives. In $k$-space this corresponds to scattering of electrons which
are near but not $\it on$ the Fermi surface.
  
Note that the coupling constants $g_{i}^{\prime }$ has dimension of the inverse of energy.
Therefore each term of Eq. (\ref{hXgeo}) has
dimension 3, they are irrelevant. As only these irrelevant operators appear
in the low energy fermionic field theory of the correlated hopping term $H_{X}$, 
one might be tempted to conclude that there is no contribution of
these terms to the physical behavior of the system, in contrast to the
numerical results discussed in Section \ref{1p2}. \ How could we account of
this situation with our field theoretical analysis? In fact, 
from a renormalization group (RG) point of view, all the operators allowed by
symmetry should be included in the effective theory. The fact that operators
present in Eq. (\ref{hhubgeo}) were not obtained in the derivation of Eq. (\ref{hXgeo}) 
means that in the initial conditions, the different $g$ do not
depend on $X$. However, they could acquire an $X$ dependence by the
couplings between $g$ and $g^{\prime }$ when the RG flow evolves.

We can take advatage of the well known studies of thermal critical phenomena with 
RG \cite{Amit} to further understand this issue.
For this case we know that the
irrelevance of an operator means that the critical exponents are not affected 
by its presence in the Hamiltonian. 
However the critical temperature does depend on this operator. 
In our case the presence of irrelevant operators will be crucial to determine
the boundaries in parameter space of the different phases, 
where the spin or charge gap opens. 

Finally, we note that the SU(2) invariance of the ordinary Hubbard model implies that under RG
flow, $g_{2\perp }=g_{1\perp }$ remains \cite{giama}. Similarly, it is shown
in \ref{a1} that $g_{2\perp }^{\prime }=g_{1\perp }^{\prime }$ is
required to keep the SU(2) invariance of the full Hamiltonian.

\subsection{Bosonization}

\label{2p2}
Bosonization is a powerful technique to analyze interacting one-dimensional 
fermionic systems \cite{giama}. Some of the interacting terms in the fermionic 
Hamiltonian become free non-interacting
terms in the bosonic Hamiltonian. The remaining terms contain in
general cosines of the bosonic fields.
Their effect can be studied by a perturbative implementation of the RG method.
If in the RG flow the coefficient of a cosine decreases, the fixed point corresponds
to a trivial theory of free bosons with known properties.  
When the RG flow goes to strong coupling the coefficient of a cosine increases. The fields are trapped in a minimum of the free energy
and the different phases can be characterized by 
calculating the classical value at this minimum of the bosonic operators corresponding to the physical observables.
In our case, the RG analysis is more involved, but as we shall show, it leads to a tractable theory
and correct results.

Let us therefore resort to a bosonic representation of the fermionic theory of Eq. (\ref{hXgeo})      
We use the following bosonization formula for the left ($-$) and right ($+$)
fermions\cite{bosonreview} : 
\begin{eqnarray}
\psi _{\sigma \pm }(x) &=&\frac{F_{\sigma \pm }}{\sqrt{L}}\colon e^{\mp
i\phi _{\sigma \pm }}\colon  \label{bosno} \\
&=&\frac{F_{\sigma \pm }}{\sqrt{2\pi \alpha }}e^{\mp i\phi _{\sigma \pm }}
\label{bosnotno}
\end{eqnarray}
Equation (\ref{bosno}) is normal ordered and therefore does not contain a
somewhat uncomfortable short range cutoff $\alpha $, $F$ is the Klein factor
and $L$ the length of the chain. Eq. (\ref{bosnotno}) arises from Eq. (\ref%
{bosno}) by the explicit expansion of $\phi _{\sigma \pm }$ in term of the
boson creations ($b_{\sigma \pm }$) and annihilations ($b_{\sigma \pm
}^{\dagger }$). It is given by \cite{giama,bosonreview} 
\begin{equation}
\phi_{\sigma \pm}(x)=\mp \underbrace{ i\sum_{n_p>0}\frac{e^{\mp ipx-\alpha
p/2}}{\sqrt{n_p}}b^{\dagger}_{\sigma \pm}}_{\varphi^{\dagger}_{\sigma \pm}}
\pm \underbrace{ i\sum_{n_p>0}\frac{e^{\pm ipx-\alpha p/2}}{\sqrt{n_p}}%
b_{\sigma \pm}}_{\varphi_{\sigma \pm}},  \label{expphi}
\end{equation}
where $\varphi _{\sigma \pm }^{\dagger }$ ($\varphi _{\sigma \pm }$) are the
creation (annihilation) part of the field $\phi _{\sigma \pm }$ and $p=\frac{%
Ln_{p}}{2\pi }$. We introduce the charge and spin bosonic fields $\phi _{pc}$
and $\phi _{ps}$ ($p=+,-$) 
\begin{equation}
\phi _{pc}=\frac{(\phi _{p\uparrow }+\phi _{p\downarrow })}{\sqrt{2}}{\rm , \; 
}\phi _{ps}=\frac{(\phi _{p\uparrow }-\phi _{p\downarrow })}{\sqrt{2}}.
\label{cs}
\end{equation}%
We also introduce phase fields $\phi _{m}$ and $\theta _{m}$ ($m=\uparrow
,\downarrow ,c,s$) 
\begin{equation}
\phi _{m}=\frac{\phi _{m+}+\phi _{m-}}{2}{\rm , \; }\theta _{m}=\frac{\phi
_{m-}-\phi _{m+}}{2}  \label{thatphi}
\end{equation}

The line before the last in Eq. (\ref{hXgeo}) bosonizes as: 
\begin{eqnarray}
g_{1\perp }^{\prime }) &\sum_{\sigma }&(-\partial _{x}\psi _{\sigma
+}^{\dagger }\psi _{\sigma -}\psi _{\overline{\sigma }-}^{\dagger }\psi _{%
\overline{\sigma }+}+\partial _{x}\psi _{\sigma -}^{\dagger }\psi _{\sigma
+}\psi _{\overline{\sigma }+}^{\dagger }\psi _{\overline{\sigma }-}-{\rm %
H.c.})=  \nonumber \\
&-&\frac{i}{(2\pi \alpha )^{2}}\sum_{\sigma }[e^{2i\phi _{\sigma }}\partial
_{x}\phi _{\sigma +}e^{-2i\phi _{\overline{\sigma }}}+e^{-2i\phi _{\sigma
}}\partial _{x}\phi _{\sigma -}e^{2i\phi _{\overline{\sigma }}}]-{\rm H.c.}=
\nonumber \\
&-&\frac{4\sqrt{2}i}{(2\pi \alpha )^{2}}\cos (2\sqrt{2}\phi _{s})\partial
_{x}\phi _{c}=-\frac{4\sqrt{2}i}{L^{2}}\colon \cos (2\sqrt{2}\phi
_{s})\colon \partial _{x}\phi _{c}  \label{g1pbos}
\end{eqnarray}%
In the last line we have normal ordered the cosine using Eqs. (\ref{bosno}) and (\ref{bosnotno}).
We also have: 
\begin{eqnarray}
g_{3\perp }^{\prime }) &\sum_{\sigma }&[-\partial _{x}\psi _{\sigma
-}^{\dagger }\psi _{\sigma +}\psi _{\overline{\sigma }-}^{\dagger }\psi _{%
\overline{\sigma }+}+\partial _{x}\psi _{\sigma +}^{\dagger }\psi _{\sigma
-}\psi _{\overline{\sigma }+}^{\dagger }\psi _{\overline{\sigma }-}-h.c.]= 
\nonumber \\
&\frac{4\sqrt{2}i}{(2\pi \alpha )^{2}}&\cos (2\sqrt{2}\phi _{c})\partial
_{x}\phi _{c}=\frac{4\sqrt{2}i}{L^{2}}\colon \cos (2\sqrt{2}\phi _{c})\colon
\partial _{x}\phi _{c}  \label{g3pbos}
\end{eqnarray}%
The bosonization of $g_{2\perp }^{\prime }$ and $g_{4\perp }^{\prime }$
terms is a little more subtle. It is convenient to come back to the lattice
version of the derivate with respect to $x$ as is given in Eq. (\ref{discretder}). We have: 
\begin{eqnarray}
g_{2\perp }^{\prime })\sum_{\sigma }\Bigg[ &-&\frac{\psi _{\sigma
+}^{\dagger }(x+a)-\psi _{\sigma +}^{\dagger }(x)}{a}\psi _{\sigma +}\psi _{%
\overline{\sigma }-}^{\dagger }\psi _{\overline{\sigma }-}+\frac{\psi
_{\sigma -}^{\dagger }(x+a)-\psi _{\sigma -}^{\dagger }(x)}{a}\psi _{\sigma
-}\psi _{\overline{\sigma }+}^{\dagger }\psi _{\overline{\sigma }+} 
\nonumber \\
&+&\psi _{\overline{\sigma }-}^{\dagger }\psi _{\overline{\sigma }-}\psi
_{\sigma +}^{\dagger }\frac{\psi _{\sigma +}(x+a)-\psi _{\sigma +}(x)}{a}%
-\psi _{\overline{\sigma }+}^{\dagger }\psi _{\overline{\sigma }+}\psi
_{\sigma -}^{\dagger }\frac{\psi _{\sigma -}(x+a)-\psi _{\sigma -}(x)}{a}%
\Bigg]=  \nonumber \\
\frac{1}{a}\sum_{\sigma }\Bigg[\Bigg( &-&\psi _{\sigma +}^{\dagger
}(x+a)\psi _{\sigma +}(x)+\psi _{\sigma +}^{\dagger }(x)\psi _{\sigma +}(x+a)%
\Bigg)\psi _{\overline{\sigma }-}^{\dagger }\psi _{\overline{\sigma }-} 
\nonumber \\
&+&\Bigg(\psi _{\sigma -}^{\dagger }(x+a)\psi _{\sigma -}(x)-\psi _{\sigma
-}^{\dagger }(x)\psi _{\sigma -}(x+a)\Bigg)\psi _{\overline{\sigma }%
+}^{\dagger }\psi _{\overline{\sigma }+}\Bigg]  \label{g2latt}
\end{eqnarray}%
We use Eq. (\ref{bosno}) to bosonize the first term into: 
\begin{eqnarray}
&\frac{1}{a}&\Bigg(\psi _{\sigma +}^{\dagger }(x)\psi _{\sigma +}(x+a)-\psi
_{\sigma +}^{\dagger }(x+a)\psi _{\sigma +}(x)\Bigg)=\frac{1}{La}\Bigg[%
\colon e^{i\phi _{\sigma +}(x)}\colon \colon e^{-i\phi _{\sigma
+}(x+a)}\colon -{\rm H.c.}\Bigg]=  \nonumber \\
&\frac{1}{La}&\Bigg[e^{-i\overbrace{(\varphi _{\sigma +}^{\dagger
}(x+a)-\varphi _{\sigma +}^{\dagger }(x))}^{\simeq a\partial _{x}\varphi
_{\sigma +}^{\dagger }+\frac{a^{2}}{2}\partial _{x}^{2}\varphi _{\sigma
+}^{\dagger }}}e^{(-i(\varphi _{\sigma +}(x+a)-\varphi _{\sigma +}(x))}\bigg(%
-\frac{Li}{2\pi a}\bigg)-{\rm H.c.}\Bigg]\simeq   \nonumber \\
&-\frac{1}{2\pi a^{2}}&\Bigg[i\bigg(1-ia\partial _{x}\varphi _{\sigma
+}^{\dagger }-i\frac{a^{2}}{2}\partial _{x}^{2}\varphi _{\sigma +}^{\dagger
}-\frac{a^{2}}{2}(\partial _{x}\varphi _{\sigma +}^{\dagger })^{2}\bigg) \nonumber \\
&\times&
\bigg(1-ia\partial _{x}\varphi _{\sigma +}-i\frac{a^{2}}{2}\partial
_{x}^{2}\varphi _{\sigma +}-\frac{a^{2}}{2}(\partial _{x}\varphi _{\sigma
+})^{2}\bigg)-{\rm H.c.}\Bigg]=  \nonumber \\
&-\frac{i}{2\pi a^{2}}&\Bigg[2\bigg(1-a^{2}\partial _{x}\varphi _{\sigma
+}^{\dagger }\partial _{x}\varphi _{\sigma +}-\frac{a^{2}}{2}(\partial
_{x}\varphi _{\sigma +})^{2}-\frac{a^{2}}{2}(\partial _{x}\varphi _{\sigma
+}^{\dagger })^{2}\bigg)\Bigg]=  \nonumber \\
&\frac{i}{\pi a^{2}}&\Bigg[-1+\frac{a^{2}}{2}\colon (\partial _{x}\phi
_{\sigma +})^{2}\colon \Bigg]  \label{term1}
\end{eqnarray}
In the second equality we have used the identity \cite{bosonreview} 
$e^{A}e^{B}=e^{B}e^{A}e^{[A,B]}$ being the commutator a c number which could
be calculated by the explicit expansion given in Eq. (\ref{expphi}) 
with $\alpha =0$. The commutator becomes: 
\begin{eqnarray}
&&\lbrack \varphi _{\sigma +}(x),\varphi _{\sigma +}^{\dagger }(x+a)]
=\sum_{p,p^{\prime }>0}\frac{e^{ipx}e^{-ip^{\prime }(x+a)}}{\sqrt{%
n_{p}n_{p}^{\prime }}}\underbrace{[b_{p},b_{p^{\prime }}^{\dagger }]}%
_{\delta _{p,p^{\prime }}} \nonumber \\
&&=\sum_{p}\frac{e^{-ipa}}{n_{p}}=-\log (1-e^{-i \frac{2\pi a}{L}}) 
\underbrace{\simeq }_{L>>a}-\log (i\frac{2\pi a}{L}). 
\label{conm0}
\end{eqnarray}
Then,
\begin{eqnarray}
&&e^{[\varphi _{\sigma +}(x),\varphi _{\sigma +}^{\dagger }(x+a)]} =-\frac{iL%
}{2\pi a},  \label{conm}
\end{eqnarray}
the value previously used.

Proceeding in a similar way one finds (to be used later)

\begin{equation}
e^{[\varphi _{\sigma -}(x),\varphi _{\sigma -}^{\dagger }(x\pm a)]}=\pm 
\frac{iL}{2\pi a}.  \label{conm2}
\end{equation}

The term into the parenthesis in the last line of (\ref{g2latt}) can be
bosonized by similar steps. The result is: 
\begin{equation}
\frac{1}{a}\Bigg(\psi _{\sigma -}^{\dagger }(x+a)\psi _{\sigma -}(x)-\psi
_{\sigma -}^{\dagger }(x)\psi _{\sigma -}(x+a)\Bigg)=\frac{i}{\pi a^{2}}%
\Bigg[-1+\frac{a^{2}}{2}\colon (\partial _{x}\phi _{\sigma -})^{2}\colon %
\Bigg]  \label{term2}
\end{equation}%

Taking into account that the normal ordered densities bozonize as: 
\[
\colon \rho _{\sigma \pm }\colon =\colon \psi _{\sigma \pm }^{\dagger }\psi
_{\sigma \pm }\colon =-\frac{1}{2\pi }\colon \partial _{x}\phi _{\sigma \pm
}\colon 
\]%
we obtain the bosonized expression of Eq. (\ref{g2latt}): 
\begin{eqnarray}
&\frac{i}{2(\pi a)^{2}}&\sum_{\sigma }\Bigg[(\partial _{x}\phi _{\sigma
+}+\partial _{x}\phi _{\sigma -})-\frac{a^{2}}{2}\Bigg((\partial _{x}\phi
_{\sigma +})^{2}\partial _{x}\phi _{\overline{\sigma }-}+(\partial _{x}\phi
_{\sigma -})^{2}\partial _{x}\phi _{\overline{\sigma }+}\Bigg)\Bigg] 
\nonumber \\
=\frac{i}{2(\pi a)^{2}} &&\Bigg[\sqrt{2}\bigg(\partial _{x}\phi
_{c+}+\partial _{x}\phi _{c-}\bigg)-  \nonumber \\
&\frac{a^{2}}{2\sqrt{2}}&\bigg((\partial _{x}\phi _{c-})^{2}\partial
_{x}\phi _{c+}+\partial _{x}\phi _{c+}\bigg((\partial _{x}\phi
_{s-})^{2}-2\partial _{x}\phi _{s-}\partial _{x}\phi _{s+}\bigg)+  \nonumber
\\
&&\partial _{x}\phi _{c-}\bigg((\partial _{x}\phi _{c+})^{2}-2\partial
_{x}\phi _{s-}\partial _{x}\phi _{s+}+(\partial _{x}\phi _{s+})^{2}\bigg)%
\Bigg)\Bigg]  \label{g2boso}
\end{eqnarray}%
Quite similar steps leads to the bosonization of the $g_{4}^{\prime }$ term.
We find: 
\begin{eqnarray}
&\frac{i}{2(\pi a)^{2}}&\sum_{\sigma }\Bigg[(\partial _{x}\phi _{\sigma
+}+\partial _{x}\phi _{\sigma -})-\frac{a^{2}}{2}\Bigg((\partial _{x}\phi
_{\sigma +})^{2}\partial _{x}\phi _{\overline{\sigma }+}+(\partial _{x}\phi
_{\sigma -})^{2}\partial _{x}\phi _{\overline{\sigma }-}\Bigg)\Bigg] 
\nonumber \\
=\frac{i}{2(\pi a)^{2}} &&\Bigg[\sqrt{2}\bigg(\partial _{x}\phi
_{c+}+\partial _{x}\phi _{c-}\bigg)-  \nonumber \\
&\frac{a^{2}}{2\sqrt{2}}&\bigg((\partial _{x}\phi _{c-})^{3}+(\partial
_{x}\phi _{c+})^{3}-\partial _{x}\phi _{c-}(\partial _{x}\phi
_{s-})^{2}-\partial _{x}\phi _{c+}(\partial _{x}\phi _{s+})^{2}\Bigg)\Bigg]
\label{g4boso}
\end{eqnarray}%
Collecting the different pieces, going to an imaginary time $\tau =it$ and
defining complex space-time coordinates ($z=v_{F}\tau +ix$,$\overline{z}%
=v_{F}\tau -ix$), where $v_{F}=a(t-X)$ is the Fermi velocity (starting from
a Hartree-Fock decoupling \cite{bos}) we obtain the following expression for
the part of the action proportional to $X$.

\begin{equation}
S_{X}=a(G_{1}^{\prime }\int d^{2}rO_{1}^{\prime }(r)+G_{3}^{\prime }\int
d^{2}rO_{3}^{\prime }{d^{2}r}+G_{2}^{\prime }\int d^{2}rO_{2}^{\prime }{%
d^{2}r}+G_{4}^{\prime }\int d^{2}rO_{4}^{\prime }{d^{2}r)}  \label{Szzvar}
\end{equation}%
where $G_{\alpha }^{\prime }=g_{\perp \alpha }^{\prime }/(a\pi v_{F})$ and $%
d^{2}r=v_{F}dxd\tau $. The different operators in Eq. (\ref{Szzvar}) are: 

\begin{eqnarray}
O_{1}^{\prime } &=&i\frac{2\pi \sqrt{2}}{L^{2}}\colon \cos (2\sqrt{2}\phi
_{s})\colon (\partial _{z}\phi _{c-}-\partial _{\overline{z}}\phi _{c+}) 
\nonumber \\
O_{3}^{\prime } &=&-i\frac{2\pi \sqrt{2}}{L^{2}}\colon \cos (2\sqrt{2}\phi
_{c})(\partial _{z}\phi _{c-}-\partial _{\overline{z}}\phi _{c+})\colon 
\nonumber \\
O_{2}^{\prime } &=&i\Bigg[\underbrace{\frac{\sqrt{2}}{2\pi a^{2}}\bigg(%
\partial _{\overline{z}}\phi _{c+}-\partial _{z}\phi _{c-}\bigg)}%
_{O_{2.1}^{\prime }}+\frac{1}{4\sqrt{2}\pi }\bigg(\underbrace{\colon
(\partial _{z}\phi _{c-})^{2}\partial _{\overline{z}}\phi _{c+}\colon }%
_{O_{2.2}^{\prime }} \nonumber \\
&+& \underbrace{\partial _{\overline{z}}\phi
_{c+}\colon (\partial _{z}\phi _{s-})^{2}\colon }_{O_{2.3}^{\prime }}+%
\underbrace{2\partial _{\overline{z}}\phi _{c+}\partial _{z}\phi
_{s-}\partial _{\overline{z}}\phi _{s+}}_{O_{2.4}^{\prime }} 
\nonumber \\
&-&\bigg(\underbrace{\partial _{z}\phi _{c-}\colon (\partial _{\overline{z}%
}\phi _{c+})^{2}\colon }_{O_{2.5}^{\prime }}+\underbrace{2\partial _{z}\phi
_{c-}\partial _{z}\phi _{s-}\partial _{\overline{z}}\phi _{s+}}%
_{O_{2.6}^{\prime }}+\underbrace{\partial _{z}\phi _{c-}\colon (\partial _{%
\overline{z}}\phi _{s+})^{2}\colon }_{O_{2.7}^{\prime }}\bigg)\Bigg)\Bigg] 
\nonumber \\
O_{4}^{\prime } &=&i\Bigg[\frac{\sqrt{2}}{2a^{2}\pi }\bigg(\partial _{%
\overline{z}}\phi _{c+}-\partial _{z}\phi _{c-}\bigg)-  \nonumber \\
&\frac{1}{4\sqrt{2}\pi }&\bigg((\partial _{z}\phi _{c-})^{3}-(\partial _{%
\overline{z}}\phi _{c+})^{3}-\partial _{z}\phi _{c-}(\partial _{z}\phi
_{s-})^{2}+\partial _{\overline{z}}\phi _{c+}(\partial _{\overline{z}}\phi
_{s+})^{2}\Bigg)\Bigg]  \label{defO'}
\end{eqnarray}

To obtain Eq. (\ref{Szzvar}) we have:

\begin{enumerate}
\item Included the normal order of each bosonic operator assuming that the
original fermionic operators were already normal ordered. This is a
prerequisite for the bosonization to work \cite{bosonreview}.

\item Taken into account that $\partial_x=i(\partial_z-\partial_{\overline {z%
}})$ and

\item that the right and left bosons depend on $\overline{z}$ and $z$
respectively. I.e. $\phi _{m+}(\overline{z})$ and $\phi _{m-}({z})$,   ($m=c$
or $s$).
\end{enumerate}

This last fact arises when the explicit time dependence of the bosonic
creation and annihilation operator is deduced from the Heisenberg equations of motion
using the free bosonic Hamiltonian $H_{0}=v_{F}\sum_{k>0}kb_{mk+}^{\dagger
}b_{mk+}+v_{F}\sum_{k>0}kb_{mk-}^{\dagger }b_{mk-}$. One obtains $b_{mk\pm
}(\tau )=\exp ^{-v_{F}\tau k}b_{mk\pm }(0)$, and $b_{mk\pm }^{\dagger }(\tau
)=\exp ^{v_{F}\tau k}b_{mk\pm }^{\dagger }(0)$. Plugging these expressions
in equations like (\ref{expphi}) one obtains: 
\begin{eqnarray}
\phi _{m+}(\overline{z}) &=&\underbrace{-i\sum_{n_{p}>0}\frac{e^{\overbrace{%
-ipx+pv_{F}\tau }^{p\overline{z}}}}{\sqrt{n_{p}}}b_{mp+}^{\dagger }}%
_{\varphi _{m+}^{\dagger }}+\underbrace{i\sum_{n_{p}>0}\frac{e^{\overbrace{%
ipx-pv_{F}\tau }^{-p\overline{z}}}}{\sqrt{n_{p}}}b_{mp+}}_{\varphi _{m+}} 
\nonumber \\
\phi _{m-}(z) &=&\underbrace{-i\sum_{n_{p}>0}\frac{e^{\overbrace{%
-ipx-pv_{F}\tau }^{-pz}}}{\sqrt{n_{p}}}b_{mp-}}_{\varphi _{m-}}+\underbrace{%
i\sum_{n_{p}>0}\frac{e^{\overbrace{ipx+pv_{F}\tau }^{pz}}}{\sqrt{n_{p}}}%
b_{mp-}^{\dagger }}_{\varphi _{m-}^{\dagger }}  \label{phizvarz}
\end{eqnarray}%
The total action is $S=S_{H}+S_{X}$, where $S_{H}$ is the usual bosonized
version of the Hubbard model of Eq. (\ref{hhubgeo}) 
\begin{eqnarray}
S_{H} &=&\frac{1}{2\pi }\int \partial _{z}\phi _{c-}\partial _{\overline{z}%
}\phi _{c+}d^{2}r+\frac{1}{2\pi }\int \partial _{z}\phi _{s-}\partial _{%
\overline{z}}\phi _{s+}d^{2}r+  \nonumber \\
&G_{1}&\int d^{2}rO_{1}(r)+G_{3}\int d^{2}rO_{3}{d^{2}r}+G_{2c}\int
d^{2}rO_{2c}{d^{2}r}+G_{2s}\int d^{2}rO_{2s}{d^{2}r}+  \nonumber \\
&&\delta v_{s}\int d^{2}rO_{4s}{d^{2}r}+\delta v_{c}\int d^{2}rO_{4c}{d^{2}r}
\label{Szzvarg}
\end{eqnarray}%
with
\begin{eqnarray}
O_{1} &=&\frac{2\pi }{L^{2}}\colon \cos (2\sqrt{2}\phi _{s})\colon 
\label{o1} \\
O_{3} &=&\frac{2\pi }{L^{2}}\colon \cos (2\sqrt{2}\phi _{c})\colon 
\label{o3} \\
O_{2c} &=&\frac{1}{4\pi }\partial _{z}\phi _{c-}\partial _{\overline{z}}\phi
_{c+}  \label{O2c} \\
O_{2s} &=&-\frac{1}{4\pi }\partial _{z}\phi _{s-}\partial _{\overline{z}%
}\phi _{s+}  \label{O2s} \\
O_{4c} &=&\frac{1}{8\pi }\bigg((\partial _{\overline{z}}\phi
_{c+})^{2}+(\partial _{{z}}\phi _{c-})^{2}\bigg)  \label{O4c} \\
O_{4s} &=&-\frac{1}{8\pi }\bigg((\partial _{\overline{z}}\phi
_{s+})^{2}+(\partial _{{z}}\phi _{s-})^{2}\bigg)  \label{O4s}
\end{eqnarray}
$\delta v_{c}$ ($\delta v_{s}$) renormalize the charge (spin) velocity.
Operators $O_{2c}$ and $O_{2s}$ (and $O_{4c}$ and $O_{4s}$) appear together
in the bosonized theory of the Hubbard model Eq. (\ref{hhubgeo}), where only interaction
between electron of different spin are taken into account. They are
independent operators in the general case where interaction between electron
of the same spin are included. For the Hubbard model, the values of all
couplings are $G_{\alpha }=$ $Ua/(\pi v_{F}).$

\subsection{The renormalization group equations}

\label{2p3}

Following Ref. \cite{cardy}, the RG equations for the coupling constants $%
\Gamma _{\alpha }$ ($G_{\alpha }$ or $G_{\alpha }^{\prime }$) present in the
action is

\begin{equation}
\frac{d\Gamma _{\gamma }}{dl}=(d-\Delta _{\gamma })\Gamma _{\gamma }
-\frac{S_{d}\lambda ^{\Delta _{\alpha }+\Delta _{\beta }-\Delta _{\gamma }-d}}{2}
\sum\limits_{\alpha \beta }C_{\alpha \beta }^{\gamma
}\Gamma _{\alpha }\Gamma _{\beta },  \label{rgcar}
\end{equation}%
where $d=2$ is the spacetime dimension of the system, $\Delta _{\gamma }$ is the
scaling dimension of the operator related with $\Gamma _{\gamma }$, $S_{d}$
is the area of a sphere of unit radius in $d$ dimensions ($2\pi $ in our
case),  the $C_{\alpha \beta }^{\gamma }$ are the coefficients
of the following short-distance Operator Product Expansion (OPE): 
\begin{equation}
O_{\alpha }^{\prime }(r)O_{\beta }^{\prime }(r^{\prime })=\sum_{\gamma
}C_{\alpha \beta }^{\gamma }\frac{O_{\gamma }(\frac{r+r^{\prime }}{2})}{\mid
r-r^{\prime }\mid ^{\Delta _{\alpha }+\Delta _{\beta }-\Delta _{\gamma }}}+%
\mbox{more
irrelevant operators,}  \label{opec}
\end{equation}
 and $\lambda$ is a number of order one, which comes from our definition of the short distance
cutoff as $a/\lambda$ (the exponent of $\lambda$ in Eq. (\ref{rgcar}) comes from the integral of
Eq. (\ref{opec}) with respect to $\mid r-r^{\prime }\mid$ in $d$ space-time dimensions).

The OPE's between two $O_{\alpha }$ operators are already known from the RG
equations of the ordinary Hubbard model \cite{giama}. The OPE's between one 
$O_{\alpha }$ and one $O_{\beta }^{\prime }$ operator give another $O_{\gamma
}^{\prime }$ and have a prefactor $G_{\alpha }G_{\beta }^{\prime }\sim UX$.
We note that for small $X$, this product is of order $X^{3}$ on the spin
transition and of higher order or negligible on the charge transition. We
have neglected these OPE's. This is partially justified by that fact that
they generate $O_{\gamma }^{\prime }$ operators which are irrelevant, while
as we show below the OPE's between two $O_{\alpha }^{\prime }$ operators
generate marginal $O_{\alpha }$ operators. A deeper justification in given
on symmetry grounds: expressing the first Eq. (\ref{sym}) in terms of the
Hartree-Fock hopping $\tilde{t}=t-X$ [which is invariant under the
transformation $c_{i\sigma }^{\dagger }\rightarrow (-1)^{i}c_{i\sigma }$]
one has

\begin{equation}
H(\tilde{t},-X,U)\equiv H(\tilde{t},X,U).  \label{sym2}
\end{equation}%
This means that for small $X$, there can be no terms of order $UX$ in the
action which correct the Hartree-Fock results. Therefore, the generated
operators in the OPE's between one $O$ and one 
$O_{\alpha }^{\prime }$ should introduce corrections of higher order.

Now let us discuss the different operators that could arise from the OPE's
between two $O_{\alpha }^{\prime }$ operators. There are some cases where
these OPE's give operators of dimension $4$ or higher. This is for example
the case of the OPE between $O_{2.3}$ and $O_{2.7}$ or in general between
two operators included in $O_{2}$   
which contain less than two fields in common.
There are other cases where the denominator does not depend only on the
distance between the two points under consideration but have factors of the
form $(z^{\prime }-z)^{-2}+(\overline{z}^{\prime }-\overline{z})^{-2}$. This
gives rise to a periodic function in the relative angle and the integral in
the angular part of $r-r^{\prime }$ [which was performed to arrive at 
Eq. (\ref{rgcar})] vanishes. This is the cases of the OPE between $O_{1}^{\prime }$
and $(O_{2.4}^{\prime }+O_{2.6}^{\prime })$. Finally there are cases which
produce operators of the form (\ref{O2c}) and (\ref{O2s}). They simply
renormalize the charge or spin velocity and will not be taken into account
in our treatment 
The remaining OPE's are displayed in the \ref{OPESO}. From this appendix we have: 
\begin{eqnarray}
C_{2^{\prime }1^{\prime }}^{1} &=&C_{1^{\prime }2^{\prime }}^{1}=\frac{1}{%
2\pi }{\rm , \; }C_{2^{\prime }3^{\prime }}^{3}=C_{3^{\prime }2^{\prime
}}^{3}=-\frac{1}{2\pi }  \nonumber \\
C_{2^{\prime }2^{\prime }}^{2s} &=&C_{2^{\prime }2^{\prime }}^{2c}=\frac{1}{%
\pi }.  \label{Gs}
\end{eqnarray}

From Eqs. (\ref{rgcar}) and (\ref{Gs}), the usual RG equations for the
Hubbard model become modified as follows. For the charge sector

\begin{equation}
\frac{dG_{2c}}{dl}=G_{3}^{2}-\lambda ^{2}(G_{2}^{\prime })^{2}{\rm , \; }\frac{%
dG_{3}}{dl}=G_{2c}G_{3}+\lambda ^{2}G_{2}^{\prime }G_{3}^{\prime },
\label{ch}
\end{equation}%
for the spin sector

\begin{equation}
\frac{dG_{2s}}{dl}=-G_{1}^{2}-\lambda ^{2}(G_{2}^{\prime })^{2}{\rm , \; }%
\frac{dG_{1}}{dl}=-G_{2s}G_{1}-\lambda ^{2}G_{2}^{\prime }G_{1}^{\prime },
\label{sp}
\end{equation}%
and in addition

\begin{equation}
\frac{dG_{\alpha }^{\prime }}{dl}=-G_{\alpha }^{\prime }.  \label{gp}
\end{equation}

\subsection{Analysis of the RG equations}

\label{2p4}

Taking into account the initial conditions, Eq. (\ref{gp}) can be integrated
immediately giving
\begin{equation}
G_{\alpha }^{\prime }=\frac{aX}{\pi v_{F}}\exp (-l).  \label{gp2}
\end{equation}%

Replacing this equation in Eqs. (\ref{ch}), one obtains two coupled
differential equations for the charge sector

\begin{equation}
\frac{dG_{2c}}{dl}=G_{3}^{2}-Ae^{-2l}{\rm , \; }\frac{dG_{3}}{dl}%
=G_{2c}G_{3}+Ae^{-2l}{\rm , \; with \;}A=\left( \frac{\lambda aX}
{\pi v_{F}}\right)^{2},  \label{ch2}
\end{equation}
with the initial conditions $G_{2c}(l=0)=G_{3}(l=0)=Ua/(\pi v_{F})$. 

It is known that for $A=0$, the flux continues along the separatrix 
$G_{2c}=G_{3}$, and goes to infinite $G_{i}$ (charge gap) if $U>0$, 
and to $G_{2c}=G_{3}=0$ (gapless case) if $U<0$. While an analytical solution for 
$A\neq 0$ seems not possible, it is clear that the effect of $A$ is to push
the flux perpendicularly to the separatrix, favoring larger 
$G_3$ and smaller $G_{2c}$. This does not modify the
final result that the critical value of $U$ which separates the regions of
diverging or vanishing $G_{3}(l\rightarrow +\infty )$ is $U_{c}=0$. We have
confirmed this by a numerical study of Eqs. (\ref{ch2}).
However, as a difference with the Hubbard model for which the flux is on the
separatrix, in our case, for $U<U_c$ (when $G_3$ flows to zero), 
$G_{2c}$ converges to a negative value. This leads to a correlation exponent \cite{giama}
$K_c \sim 1- G_{2c}$ larger than 1. As a consequence, the singlet superconducting (SS)
correlation functions, which decay as $d^{-1/K_c}$ at large distance $d$ dominate over the
charge density wave (CDW) ones, which decay as $d^{-K_c}$ \cite{jaka,bos,giama}.
In the Hubbard model, for $U<0$, $K_c =1$ and both SS and CDW correlations decay as $1/d$.

For the spin sector, the RG equations become

\begin{equation}
\frac{dG_{2s}}{dl}=-G_{1}^{2}-Ae^{-2l}{\rm , \; }\frac{dG_{1}}{dl}%
=-G_{1}G_{2s}-Ae^{-2l},.  \label{rgs1}
\end{equation}%
with the initial conditions $G_{2s}(0)=G_{1}(0)=Ua/(\pi v_{F})$. 

It is clear that the flux of the RG equations remains on the
separatrix $G_{1}(l)=G_{2s}(l)$. Therefore, both equations (\ref{rgs1})
reduce to the same equation for $G_{1}=G_{2s}=G$. Changing variable 
$z=\sqrt{A}e^{-l}$, this equation takes the form

\begin{equation}
z\frac{dG}{dz}=G^{2}+z^{2}.  \label{rgs2}
\end{equation}%
Its solution is given in terms of Bessel functions

\begin{equation}
G(z)=\frac{z[Y_{1}(z)+CJ_{1}(z)]}{Y_{0}(z)+CJ_{0}(z)},  \label{rgsol}
\end{equation}%
where the constant $C$ is determined by the initial condition $G(\sqrt{A})=G_{\rm ini}$, giving

\begin{equation}
C=\frac{G_{\rm ini}Y_{0}(\sqrt{A})-\sqrt{A}Y_{1}(\sqrt{A})}{-G_{\rm ini}J_{0}(\sqrt{A})+%
\sqrt{A}J_{1}(\sqrt{A})}.  \label{c}
\end{equation}

Mathematically, for $l\rightarrow \infty $ $(z\rightarrow 0)$, 
Eq. (\ref{rgsol}) converges to zero. However, it may happen that $G(z_{d})$ diverges
for some intermediate value $z_{d}$ ($0<z<\sqrt{A}$),
jumping from $-\infty$ to $+\infty$ as $z$ decreases. 
This means
physically that at an intermediate scale determined by $z_d$, 
the solution flowed to the strong coupling fixed point at which a
spin gap opens. 
The limiting value of $z_{d}$ for which such a behavior
takes place corresponds to $z_{d}\rightarrow 0$. Since for small values of
the argument $Y_{0}(z)\sim (2/\pi )\ln (z)$ and $J_{0}(z)\sim 1$, 
a diverging $G(z_d)$ for $z_d \rightarrow 0$ implies a zero in the denominator of Eq. (\ref{rgsol}), and the
initial conditions should be such that $C$ also diverges
in this special case.

For small values of $A$
(as we have assumed in our whole treatment), there is no divergence in $G(z)$
if $C$ is negative. From this reasoning and Eq. (\ref{c}), we obtain the
following condition for the opening of a spin gap:

\begin{equation}
G_{\rm ini}<\frac{\sqrt{A}J_{1}(\sqrt{A})}{J_{0}(\sqrt{A})}\simeq \frac{A}{2}+\frac{%
A^{2}}{16},  \label{gl}
\end{equation}%
where the last member was obtained from a series expansion of the Bessel
functions.

From Eq. (\ref{gl}) and using $G_{\rm ini}=Ua/(\pi v_{F})$, we obtain the following critical
value of $U$ for the opening of the spin gap

\begin{equation}
U_{s}=\pi v_{F}\frac{\sqrt{A}J_{1}(\sqrt{A})}{J_{0}(\sqrt{A})}.  \label{usrg}
\end{equation}

or approximately

\begin{equation}
U_{s}\simeq \frac{(\lambda aX)^{2}}{2\pi v_{F}}+\frac{(\lambda aX)^{4}}{8(\pi v_{F})^{3}}.  
\label{usrgap}
\end{equation}

The final path taken by the RG flow
in each sector, determine the nature of the resulting phases.
As discussed above, for $U<U_c$, SS correlations dominate. For $U>U_s$ spin-spin correlations are
the largest at large distances as in the usual Hubbard model \cite{jaka,bos,giama}.

The phase in between, for $U_c<U<U_s$ is characterized by the presence of both gaps, and the 
RG flow in each sector leads to
$G_{3} \rightarrow +\infty $ and $G_{1} \rightarrow -\infty $.
To minimize the respective cosine terms in the action [See Eqs. (\ref{Szzvarg}), (\ref{o1}) and (\ref{o3})], 
the fields are frozen at the values $2\sqrt{2}\phi _{c}=\pi$ and $2\sqrt{2}\phi _{s}=0$.
As a consequence, the system has a spontaneously dimerized bond-ordering-wave (BOW) phase
with long range order. The order parameter
which takes a finite value on this phase is \cite{jaka,bos}

\begin{equation}
O_{BOW} = \sum_{i \sigma}(-1)^i (c_{i+1,\sigma }^{\dagger }c_{i\sigma }+{\rm H.c.})
\sim \sin(\sqrt{2}\phi _{c}) \cos(\sqrt{2}\phi _{s}). 
\label{bow}
\end{equation}

\section{Comparison with the numerical results}

\label{s3}

The field theoretical result for the charge transition $U_{c}=0$ obtained in
the previous Section, agrees with the numerical results, presented in
Section \ref{1p2}. As explained in Section \ref{1p3}, this result is not
obtained if the initial values of the couplings of the Hubbard model   ($%
G_{2c}(0)$ and $G_{3}(0)$) are corrected by vertex corrections in second
order in $X$ before bosonizing. We do not have a physical explanation for
this.

\begin{figure}[tbp]
\includegraphics[width=14cm]{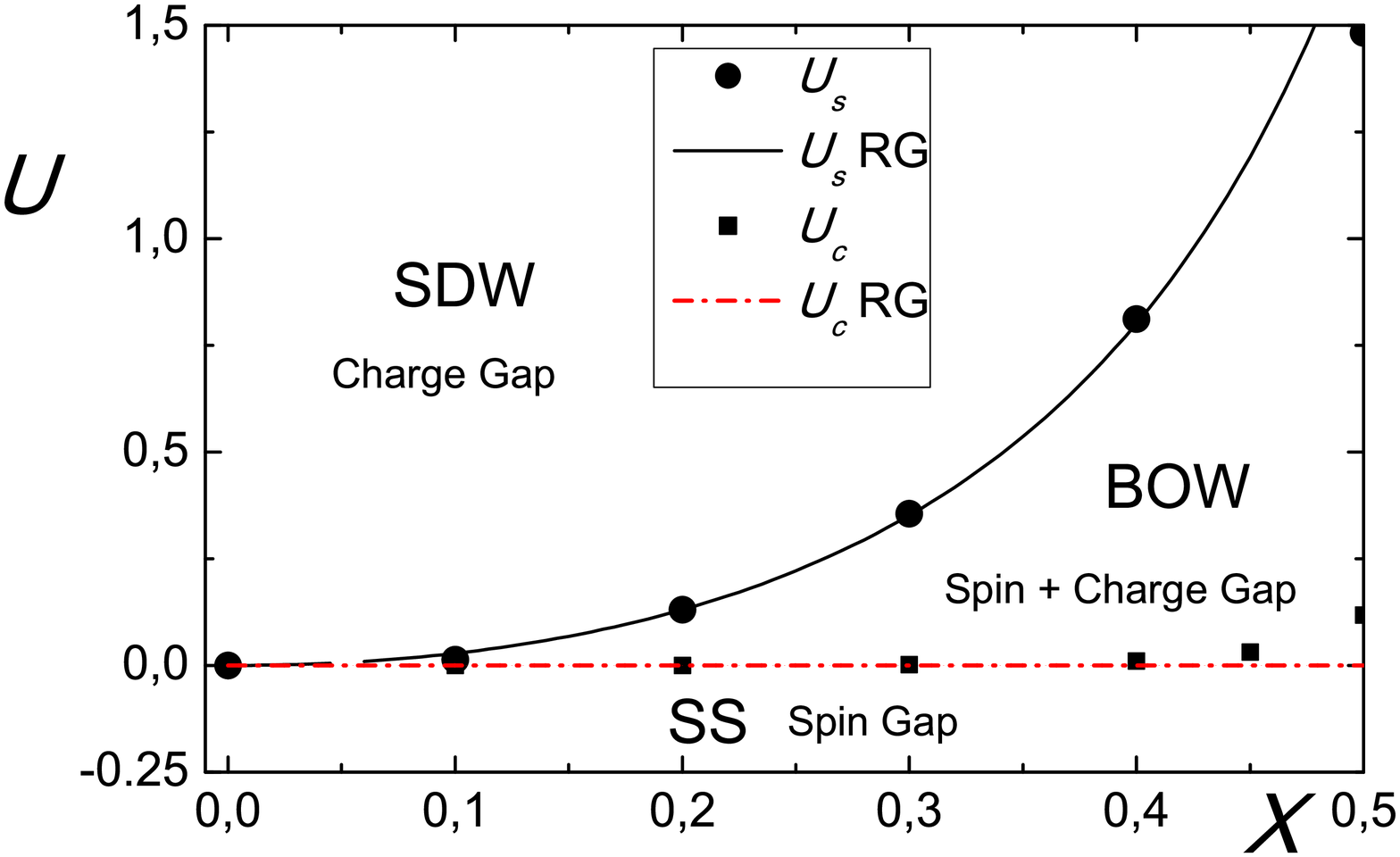}
\caption{Solid circles (squares): critical value of the spin (charge) transition $U_s$ ($U_c$) as a function of $X$
obtained from the method of level crossings. Full line: solution of the RG equations given by 
Eq. (\ref{usrg}). The dash dotted line at $U_c=0$ signal the boundary between the gapped and gapless charge phase 
as given by Eq. (\ref{ch2}). The unit of energy is taken as $t=1$.}
\label{usf}
\end{figure}

To compare the critical value of $U$ for the spin transition, we take 
$\lambda =4$, which using Eq. (\ref{usrgap}) and recalling that $v_{F}=a(t-X)$%
, leads to the same result as that obtained from vertex corrections Eq. (\ref%
{usve}) for small $X$. This leads to $A=\{4X/[\pi (t-X)]\}^{2}$. Replacing
this result in Eq. (\ref{usrg}), we obtain the function $U_{s}(X)$ that is
represented in Fig. \ref{usf}. The agreement with the numerical result up to 
$X/t\sim 0.4$ is excellent.

\section{Summary and discussion}

\label{s4}

We have studied a field theory for the Hubbard model with small bond-charge
interaction $X$ at half filling. 
While usually, it is enough to consider in the action only 
terms linear in the lattice parameter $a$, in our case it is necessary 
to include terms of order $a^2$
to obtain meaningful results at half
filling. These terms can be classified in a similar way as the linear
ones in terms of different processes in a "$g$-ology" treatment (forward one
branch, forward two branch, backward and Umklapp) but contain derivatives of
the fields in the space direction. 

We have obtained the RG equations of the different couplings using Operator
Product Expansions. While the treatment of the new terms is awkward, most of
them should be retained to keep the spin SU(2) invariance of the model.

According to the dominant correlations at large distances, the phases of the
model can be classified as a singlet superconducting (SS) one for $U<U_{c}$,
a bond ordering wave (BOW) for $U_{c}<U<U_{s}$ and a spin density wave (SDW)
for $U>U_{s}$. The boundaries between the phases correspond to a charge
transition for on site repulsion $U=U_{c}$ and a spin transition at $U=U_{s}$%
, which correspond respectively to the opening of a charge gap and a closing
of the spin gap as $U$ increases. For the former transition we obtain 
$U_{c}=0$ in agreement with previous numerical studies for $X<0.5$ \cite{cola}. 
With only one adjustable parameter, we also obtain a very good agreement
with the numerical results for $U_{s}$ if $X\lesssim 0.4t.$

To explain accurately the dependence of $U_c$ for $X>0.5$, 
it is necessary to go beyond our approach, possibly including more irrelevant operators.

As stated in Section \ref{1p1}, the model is an effective one-band model for
a variety of physical systems, in particular optical lattices \cite%
{duan,good,kest,duan2}. In these systems, $U$ can be varied over the whole
range, including its sign, through tuning of the external magnetic field $B$%
. It is also possible to change $X/t$ by 20\%. Therefore, adjusting the
filling to one particle per site, it seems in principle possible to tune the
parameters in such a way the   ground state of the system is in any of the
three phases: SS, BOW or SDW.

\section*{Acknowledgments}

We thank Pascal Simon and Luming Guan for useful discussions. We are
partially supported by CONICET, Argentina. This work was partially supported
by PIP 11220080101821 and 11220090100392 of CONICET, and PICT 2006/483, PICT 1647 and PICT R1776 of the ANPCyT.

\appendix

\section{SU(2) spin rotational symmetry}

\label{a1}

In this appendix, we derive the relations among the constants $g^{\prime }$
which leave the theory invariant under spin SU(2) transformations.

Under an SU(2) rotation each spinor $\psi _{p\sigma }$, ($p=+,-$) transform
as $\psi _{\sigma ^{\prime }}^{\prime }=U_{\sigma ^{\prime }\sigma }\psi
_{\sigma }$ (the sum over repeated indices is implicitly understood). $\mathbf{U}$ is an SU(2) matrix given by:

\begin{equation}
\mathbf{U}=\left( 
\begin{array}{cc}
a & b \\ 
-b^{\ast } & a^{\ast }%
\end{array}%
\right)   \label{su2}
\end{equation}%
$a$ and $b$ are complex number satisfying $\mid a\mid ^{2}+\mid b\mid ^{2}=1$%
. We require the invariance of the $g$-ology Hamiltonian (\ref{hXgeo}) under
this rotation. We start by undertaking a rotation of the sum of the first
terms of the second and the third line of (\ref{hXgeo}). The transformed
expression is: 
\begin{eqnarray}
&-&U_{\sigma _{1}\sigma }^{\ast }U_{\sigma \sigma _{2}}U_{\sigma _{3}%
\overline{\sigma }}^{\ast }U_{\overline{\sigma }\sigma _{4}}(\partial
_{x}\psi _{\sigma _{1}+}^{\dagger }\psi _{\sigma _{2}+}\psi _{{\sigma _{3}}%
-}^{\dagger }\psi _{{\sigma _{4}}-}+\partial _{x}\psi _{\sigma
_{1}+}^{\dagger }\psi _{\sigma _{2}-}\psi _{{\sigma _{3}}-}^{\dagger }\psi _{%
{\sigma _{4}}+})=  \nonumber \\
&-&U_{\sigma _{1}\sigma }^{\ast }U_{\sigma \sigma _{2}}U_{\sigma _{3}%
\overline{\sigma }}^{\ast }U_{\overline{\sigma }\sigma _{4}}\left( \partial
_{x}\psi _{\sigma _{1}+}^{\dagger }\psi _{\sigma _{2}+}\psi _{{\sigma _{3}}%
-}^{\dagger }\psi _{{\sigma _{4}}-}-\partial _{x}\psi _{\sigma
_{1}+}^{\dagger }\psi _{\sigma _{4}+}\psi _{\sigma _{3}-}^{\dagger }\psi
_{\sigma _{2}-}+\delta _{\sigma _{2}\sigma _{3}}\psi _{\sigma
_{1}+}^{\dagger }\psi _{\sigma _{4}+}\right) =  \nonumber \\
&-&\left( U_{\sigma _{1}\sigma }^{\ast }U_{\sigma \sigma _{2}}U_{\sigma _{3}%
\overline{\sigma }}^{\ast }U_{\overline{\sigma }\sigma _{4}}-U_{\sigma
_{1}\sigma }^{\ast }U_{\sigma \sigma _{4}}U_{\sigma _{3}\overline{\sigma }%
}^{\ast }U_{\overline{\sigma }\sigma _{2}}\right) \partial _{x}\psi _{\sigma
_{1}+}^{\dagger }\psi _{\sigma _{2}+}\psi _{{\sigma _{3}}-}^{\dagger }\psi _{%
{\sigma _{4}}-} \nonumber \\
&+& U_{\sigma _{1}\sigma }^{\ast }\underbrace{U_{\sigma \sigma
_{2}}U_{\sigma _{2}\overline{\sigma }}^{\ast }}_{=\delta _{\sigma \overline{%
\sigma }}=0}U_{\overline{\sigma }\sigma _{4}}\psi _{\sigma _{1}+}^{\dagger
}\psi _{\sigma _{4}+}  \label{g2g1rot}
\end{eqnarray}%
In the second term of the last line we have interchanged the dummy indices $%
\sigma _{2}$ with $\sigma _{4}$. Now it is possible to evaluate the
difference of the products of {\bf U} matrices involved in (\ref{g2g1rot}). 
We have: 
\begin{eqnarray}
U_{\sigma _{1}\sigma }^{\ast }U_{\sigma \sigma _{2}}U_{\sigma _{3}\overline{%
\sigma }}^{\ast }U_{\overline{\sigma }\sigma _{4}}-U_{\sigma _{1}\sigma
}^{\ast }U_{\sigma \sigma _{4}}U_{\sigma _{3}\overline{\sigma }}^{\ast }U_{%
\overline{\sigma }\sigma _{2}} &=&U_{\sigma _{1}\sigma }^{\ast }U_{\sigma
_{3}\overline{\sigma }}^{\ast }(\delta _{\sigma _{2}\overline{\sigma }%
_{4}}(\delta _{\sigma \sigma _{2}}-\delta _{\overline{\sigma }\sigma _{2}}))=
\nonumber \\
\delta _{\sigma _{2}\overline{\sigma }_{4}}(U_{\sigma _{1}\sigma _{2}}^{\ast
}U_{\sigma _{3}\overline{\sigma }_{2}}^{\ast }-U_{\sigma _{1}\overline{%
\sigma }_{2}}^{\ast }U_{\sigma _{3}{\sigma _{2}}}^{\ast }) &=&\delta
_{\sigma _{2}\overline{\sigma }_{4}}\delta _{\sigma _{3}\overline{\sigma }%
_{1}}(\delta _{\sigma _{1}\sigma _{2}}-\delta _{\sigma _{2}\sigma _{3}})
\label{Usum}
\end{eqnarray}%
Inserting in (\ref{g2g1rot}) we obtain: 
\begin{eqnarray}
(\ref{g2g1rot}) &=&-(\partial _{x}\psi _{\sigma +}^{\dagger }\psi _{\sigma
_{+}}\psi _{\overline{\sigma }-}^{\dagger }\psi _{\overline{\sigma }%
-}-\partial _{x}\psi _{\sigma +}^{\dagger }\psi _{\overline{\sigma }%
_{+}}\psi _{\overline{\sigma }-}^{\dagger }\psi _{{\sigma }-})=  \nonumber \\
&&-(\partial _{x}\psi _{\sigma +}^{\dagger }\psi _{\sigma _{+}}\psi _{%
\overline{\sigma }-}^{\dagger }\psi _{\overline{\sigma }-}+\partial _{x}\psi
_{\sigma +}^{\dagger }\psi _{\sigma -}\psi _{\overline{\sigma }-}^{\dagger
}\psi _{\overline{\sigma }+})  \label{u22}
\end{eqnarray}%
This is the sum of the first terms with the coefficients $g_{1\perp
}^{\prime }$ and $g_{2\perp }^{\prime }$. Therefore we have shown that the
sum of these terms is invariant under an SU(2) rotation. With the same
procedure we can show indeed that the sum of all the terms multiplying 
$g_{1\perp }^{\prime }$ and the ones multiplying $g_{2\perp }^{\prime }$
transform into themselves by an SU(2) rotation. This implies that under the
condition $g_{1\perp }^{\prime }=g_{2\perp }^{\prime }$, the Hamiltonian $%
H_{X}$ of Eq. (\ref{hXgeo}) remains invariant.

\section{OPE's between $O^{\prime}$ operators}

\label{OPESO} 

In this appendix we give the expressions for all non-vanishing OPE's between any two $O^{\prime}$ operators. 
The first one is: 
\begin{eqnarray}
&&{O_{2.7}^{\prime }(z,\overline{z})}.{O_{1}^{\prime }(z^{\prime },\overline{%
z^{\prime }})} =\frac{1}{4\sqrt{2}\pi }\colon \partial _{z}\phi
_{c-}(z)(\partial _{\overline{z}}\phi _{s+}(\overline{z}))^{2}\colon \nonumber \\
&\times&
\frac{2\pi \sqrt{2}}{L^{2}}\colon \cos (2\sqrt{2}\phi _{s}(z^{\prime },%
\overline{z}^{\prime })(\partial _{z^{\prime }}\phi _{c-}(z^{\prime
})-\partial _{\overline{z}^{\prime }}\phi _{c+}(\overline{z}^{\prime
}))\colon =  \nonumber \\
&&\frac{1}{2L^{2}}\bigg(\colon \cos (2\sqrt{2}\phi _{s}(z^{\prime },%
\overline{z}^{\prime }))(\partial _{\overline{z}^{\prime }}\phi
_{s+})^{2}\colon +\frac{2\sqrt{2}}{\overline{z}^{\prime }-\overline{z}}%
\colon \sin (2\sqrt{2}\phi _{s+}(z^{\prime },\overline{z}^{\prime
}))\partial _{\overline{z}}\phi _{s+}\colon   \nonumber \\
&&+\frac{2}{(\overline{z}^{\prime }-\overline{z})^{2}}\colon \cos (2\sqrt{2}%
\phi _{s}(z^{\prime },\overline{z}^{\prime }))\colon \bigg)\bigg(-\frac{1}{({%
z}^{\prime }-{z})^{2}}+\colon \partial _{z}\phi _{c-}\partial _{z^{\prime
}}\phi _{c-}\colon -\partial _{z}\phi _{c-}\partial _{\overline{z}^{\prime
}}\phi _{c+}\bigg)\simeq   \nonumber \\
&&\bigg(\frac{1}{L^{2}}\bigg)\frac{\colon \cos (2\sqrt{2}\phi _{s}(z,%
\overline{z}))\colon }{\mid {z}^{\prime }-{z}\mid ^{4}}+...=\frac{1}{2\pi }%
\frac{O_{1}}{\mid {z}^{\prime }-{z}\mid ^{4}}+...  \label{ope271}
\end{eqnarray}%
To obtain the previous result we have first normal ordered each product of
operators containing fields at different points. Then we have expanded the
resulting expressions for $z^{\prime }$ near $z$. Let us show how we have
proceeded step by step : 
\begin{eqnarray}
\partial _{z}\phi _{c-}(z)\;\;\partial _{z}\phi _{c-}(z^{\prime })
&=&(\partial _{z}\varphi _{c-}^{\dagger }(z)+\partial _{z}\varphi
_{c-}(z))(\partial _{z^{\prime }}\varphi _{c-}^{\dagger }(z^{\prime
})+\partial _{z^{\prime }}\varphi _{c-}(z^{\prime }))=  \nonumber \\
&&\partial _{z}\varphi _{c-}(z)\partial _{z^{\prime }}\varphi _{c-}({%
z^{\prime }})+\partial _{z}\varphi _{c-}^{\dagger }(z)\partial _{z^{\prime
}}\varphi _{c-}^{\dagger }({z^{\prime }})+\partial _{z}\varphi
_{c-}^{\dagger }(z)\partial _{z^{\prime }}\varphi _{c-}({z^{\prime }})+ 
\nonumber \\
&&\partial _{z^{\prime }}\varphi _{c-}^{\dagger }({z^{\prime }})\partial
_{z}\varphi _{c-}(z)+\partial _{z}\partial _{z^{\prime }}[\varphi
_{c-}(z),\varphi _{c-}^{\dagger }({z^{\prime }})]  \label{opeder}
\end{eqnarray}%
and: 
\begin{eqnarray}
\lbrack \varphi _{c-}(z),\varphi _{c-}^{\dagger }({z^{\prime }})] &=& 
\nonumber \\
\sum_{n_{p},n_{p^{\prime }}}\frac{e^{-pz}e^{p^{\prime }z^{\prime }}}{\sqrt{%
n_{p}n_{p}^{\prime }}}\underbrace{[b_{c-}^{p},b_{c-}^{p^{\prime }\dagger }]}%
_{\delta _{p,p^{\prime }}} &=&\sum_{n_{p}}\frac{e^{\frac{2\pi }{L}%
n_{p}(z^{\prime }-z)}}{n_{p}}\underbrace{=}_{\mid \underbrace{e^{\frac{2\pi 
}{L}(z^{\prime }-z)}\mid <1}_{\tau ^{\prime }<\tau }}-\log (1-e^{\frac{2\pi 
}{L}(z-z^{\prime })}) \nonumber \\
&& \underbrace{\simeq}_{L>>\mid z^{\prime } - z\mid }-\log (%
\frac{2\pi }{L}(z-z^{\prime }))  \nonumber \\
\partial _{z}\partial _{z^{\prime }}[\varphi _{c-}(z),\varphi _{c-}^{\dagger
}({z^{\prime }})] &=&-\frac{1}{(z-z^{\prime })^{2}}  \label{derzzpconm}
\end{eqnarray}%
The third equality has been taken from Ref. \cite{complexbook}. The condition $%
\tau ^{\prime }<\tau $ implies that the operator should be time ordered in
decreasing order from left to right. Including Eq. (\ref{derzzpconm}) in Eq.
(\ref{opeder}) we obtain: 
\begin{equation}
\partial _{z}\phi _{c-}(z)\;\;\partial _{z}\phi _{c-}(z^{\prime })=\colon
\partial _{z}\phi _{c-}(z)\;\;\partial _{z}\phi _{c-}(z^{\prime })\colon -%
\frac{1}{(z-z^{\prime })^{2}}
\end{equation}%
which is the result used in Eq. (\ref{ope271}). Regarding the normal
ordering of the product $\colon (\partial _{\overline{z}}\phi _{s+}(%
\overline{z}))^{2}\colon \;\;\colon \cos (2\sqrt{2}\phi _{s}(z^{\prime },%
\overline{z}^{\prime })\colon $, we use the following basic OPE: 
\begin{eqnarray}
&& \partial _{\overline{z}}\varphi _{s+}(\overline{z})e^{i\lambda \varphi
_{s}^{\dagger }(\overline{z}^{\prime })} =  \nonumber \\
&&e^{i\lambda \varphi _{s}^{\dagger }(\overline{z}^{\prime })}\partial _{%
\overline{z}}\varphi _{s+}(\overline{z})+i{\lambda }\underbrace{[\partial _{%
\overline{z}}\varphi _{s+}(\overline{z}),\underbrace{\varphi _{s}^{\dagger }(%
\overline{z}^{\prime })}_{\frac{(\varphi _{s+}^{\dagger }+\varphi
_{s-}^{\dagger })}{2}}]}_{\frac{1}{2(\overline{z}^{\prime }-\overline{z})}%
}e^{i\lambda \varphi _{s}^{\dagger }(\overline{z}^{\prime })}
\end{eqnarray}%
where we have used $Ae^{B}=e^{B}A+[A,B]e^{B}$ 
and the commutator was calculated as in Eq. (\ref{derzzpconm}). Therefore, we have: 
\begin{eqnarray}
&& \partial _{\overline{z}}\varphi _{s+}(\overline{z})\colon \cos (2\sqrt{2}%
\phi _{s}(z^{\prime },\overline{z}^{\prime }))\colon  =  \nonumber \\
&&\colon \cos (2\sqrt{2}\phi _{s}(z^{\prime },\overline{z}^{\prime
}))\partial _{\overline{z}}\varphi _{s+}(\overline{z}):-\frac{\sqrt{2}}{(%
\overline{z}^{\prime }-\overline{z})}\colon \sin (2\sqrt{2}\phi
_{s}(z^{\prime },\overline{z}^{\prime }))\colon 
\end{eqnarray}%
(the last sign changes if one permutes sin and cos) and: 
\begin{eqnarray}
&&(\partial _{\overline{z}}\varphi _{s+}(\overline{z}))^{2}\colon \cos (2%
\sqrt{2}\phi _{s}(z^{\prime },\overline{z}^{\prime }))\colon =  \nonumber \\
&&\partial _{\overline{z}}\varphi _{s+}(\overline{z})\bigg(\colon \cos (2%
\sqrt{2}\phi _{s}(z^{\prime },\overline{z}^{\prime }))\colon \partial _{%
\overline{z}}\varphi _{s+}(\overline{z})-\frac{\sqrt{2}}{(\overline{z}%
^{\prime }-\overline{z})}\colon \sin (2\sqrt{2}\phi _{s}(z^{\prime },%
\overline{z}^{\prime }))\colon \bigg)=  \nonumber \\
&&\bigg(\colon \cos (2\sqrt{2}\phi _{s}(z^{\prime },\overline{z}^{\prime
}))\colon \partial _{\overline{z}}\varphi _{s+}(\overline{z})-\frac{\sqrt{2}%
}{(\overline{z}^{\prime }-\overline{z})}\colon \sin (2\sqrt{2}\phi
_{s}(z^{\prime },\overline{z}^{\prime }))\colon \bigg)\partial _{\overline{z}%
}\varphi _{s+}(\overline{z}) \nonumber \\
&& -\frac{\sqrt{2}}{(\overline{z}^{\prime }-%
\overline{z})}\bigg(\colon \sin (2\sqrt{2}\phi _{s}(z^{\prime },\overline{z}%
^{\prime }))\colon \partial _{\overline{z}}\varphi _{s+}+  \nonumber \\
&&\frac{\sqrt{2}}{(\overline{z}^{\prime }-\overline{z})}\colon \cos (2\sqrt{2%
}\phi _{s}(z^{\prime },\overline{z}^{\prime }))\colon \bigg)=  \nonumber \\
&&\colon \cos (2\sqrt{2}\phi _{s}(z^{\prime },\overline{z}^{\prime }))\colon %
\bigg((\partial _{\overline{z}}\varphi _{s+}(\overline{z}))^{2}-\frac{2}{(%
\overline{z}^{\prime }-\overline{z})^{2}}\bigg) \nonumber \\
&& -\frac{2\sqrt{2}}{(\overline{z%
}^{\prime } - \overline{z})}\colon \sin (2\sqrt{2}\phi _{s}(z^{\prime },%
\overline{z}^{\prime }))\colon \partial _{\overline{z}}\varphi _{s+}(%
\overline{z})
\end{eqnarray}%
The required OPE is: 
\begin{eqnarray}
&& \colon (\partial _{\overline{z}}\phi _{s+}(\overline{z}))^{2}\colon
\;\;\colon \cos (2\sqrt{2}\phi _{s}(z^{\prime },\overline{z}^{\prime
})\colon  =  \nonumber \\
&& \bigg[(\partial _{\overline{z}}\varphi _{s+}(\overline{z}))^{2}+2(\partial _{%
\overline{z}}\varphi _{s+}^{\dagger }(\overline{z}))(\partial _{\overline{z}%
}\varphi _{s+}(\overline{z}))+(\partial _{\overline{z}}\varphi
_{s+}^{\dagger })^{2}\bigg]\colon \cos (2\sqrt{2}\phi _{s}(z^{\prime },%
\overline{z}^{\prime })\colon  =  \nonumber \\
&& \colon \cos (2\sqrt{2}\phi _{s}(z^{\prime },\overline{z}^{\prime
}))(\partial _{\overline{z}^{\prime }}\phi _{s+})^{2}\colon \nonumber \\
&&-\frac{2\sqrt{2}%
}{\overline{z}^{\prime }-\overline{z}}\colon \sin (2\sqrt{2}\phi
_{s+}(z^{\prime },\overline{z}^{\prime }))\colon-\frac{2}{(\overline{z}^{\prime }-%
\overline{z})^{2}}\colon \cos (2\sqrt{2}\phi _{s}(z^{\prime },\overline{z}%
^{\prime })\colon  
\end{eqnarray}%
which is the other OPE used in Eq. (\ref{ope271}).

An analogous reasoning leads to the remaining non-vanishing OPE's:

\begin{eqnarray}
&&{O_{2.3}^{\prime }(z,\overline{z})}{O_{1}^{\prime }(z^{\prime },\overline{%
z^{\prime }})} = -\frac{1}{4\sqrt{2}\pi }\colon \partial _{\overline{z}%
}\phi _{c+}({\overline{z}})(\partial _{z}\phi _{s-}(z))^{2}\colon \nonumber \\ 
&& \times
\frac{2\pi \sqrt{2}}{L^{2}}\colon \cos (2\sqrt{2}\phi _{s}(z^{\prime },\overline{z}%
^{\prime })(\partial _{z^{\prime }}\phi _{c-}(z^{\prime })-\partial _{%
\overline{z}^{\prime }}\phi _{c+}(\overline{z}))\colon =  \nonumber \\
&&-\frac{1}{2L^{2}}\bigg(\colon \cos (2\sqrt{2}\phi _{s}(z^{\prime },%
\overline{z}^{\prime }))(\partial _{z^{\prime }}\phi _{s-})^{2}\colon +\frac{%
2\sqrt{2}}{z^{\prime }-z}\colon \sin (2\sqrt{2}\phi _{s+}(z^{\prime },%
\overline{z}^{\prime }))\partial _{z}\phi _{s-}\colon  \nonumber \\
&&+\frac{2}{(z^{\prime }-z)^{2}}\colon \cos (2\sqrt{2}\phi _{s}(z^{\prime },%
\overline{z}^{\prime }))\colon \bigg) 
\bigg(+\frac{1}{(\overline{z}-\overline{%
z}^{\prime })^{2}}+\partial _{\overline{z}}\phi _{c+}\partial _{\overline{z}%
^{\prime }}\phi _{c-}-\colon \partial _{\overline{z}}\phi _{c+}\partial _{%
\overline{z}^{\prime }}\phi _{c+}\colon \bigg)\simeq  \nonumber \\
&&\bigg(\frac{1}{L^{2}}\bigg)\frac{\colon \cos (2\sqrt{2}\phi _{s}(z,%
\overline{z})}{\mid {z}^{\prime }-{z}\mid ^{4}\colon }+...=\frac{1}{2\pi }%
\frac{O_{1}}{\mid {z}^{\prime }-{z}\mid ^{4}}+... \\
&&{O_{2.2}^{\prime }(z,\overline{z})}{O_{3}^{\prime }(z^{\prime },\overline{z}%
^{\prime })} = \frac{1}{4\sqrt{2}\pi }\colon \partial _{\overline{z}%
}\phi _{c+}({\overline{z}})(\partial _{z}\phi _{c-}(z))^{2}\colon  \nonumber \\ 
&& \times
\frac{2\pi \sqrt{2}}{L^{2}}\colon \cos (2\sqrt{2}\phi _{c}(z^{\prime },%
\overline{z}^{\prime }))(\partial _{z^{\prime }}\phi _{c-}(z^{\prime
})-\partial _{\overline{z}^{\prime }}\phi _{c+}(\overline{z}^{\prime
}))\colon \simeq  \nonumber \\
&&-\bigg(\frac{1}{L^{2}}\bigg)\frac{\colon \cos (2\sqrt{2}\phi _{c}(z,%
\overline{z}))\colon }{\mid {z}^{\prime }-{z}\mid ^{4}}+...=-\frac{1}{2\pi }%
\frac{O_{3}}{\mid {z}^{\prime }-{z}\mid ^{4}}+... \\
&& {O_{2.5}^{\prime }(z,\overline{z})}{O_{3}^{\prime }(z^{\prime },\overline{z}%
^{\prime })} = - \frac{1}{4\sqrt{2}\pi }\colon \partial _{{z}}\phi
_{c-}({z})(\partial _{{\overline{z}}}\phi _{c+}({\overline{z}}))^{2}\colon %
\nonumber \\ 
&& \times
\frac{2\pi \sqrt{2}}{L^{2}}\colon \cos (2\sqrt{2}\phi
_{c}(z^{\prime },\overline{z}^{\prime }))(\partial _{z^{\prime }}\phi
_{c-}(z^{\prime })-\partial _{\overline{z}^{\prime }}\phi _{c+}(\overline{z}%
^{\prime }))\colon =  \nonumber \\
&&-\bigg(\frac{1}{2L^{2}}\bigg)\partial _{z}\phi _{c-}\bigg(\colon \cos (2%
\sqrt{2}\phi _{c}(z^{\prime },\overline{z}^{\prime }))(\partial _{\overline{z%
}}\phi _{c+})^{2}\partial _{{z}}\phi _{c-}\colon -\frac{2\sqrt{2}}{\overline{%
z^{\prime }}-\overline{z}}\colon \sin (2\sqrt{2}\phi _{c}(z^{\prime },%
\overline{z}^{\prime }))\partial _{\overline{z}}\phi _{c+}\partial
_{z^{\prime }}\phi _{c-}\colon   \nonumber \\
&& - \frac{2}{(\overline{z^{\prime }}-\overline{z})^{2}}\colon \cos (2\sqrt{2}%
\phi _{c}(z^{\prime },\overline{z}^{\prime }))\partial _{z^{\prime }}\phi
_{c-}\colon \bigg)\simeq  \nonumber \\
&&-\bigg(\frac{1}{L^{2}}\bigg)\frac{\colon \cos (2\sqrt{2}\phi _{c}(z,%
\overline{z}))\colon }{\mid {z}^{\prime }-{z}\mid ^{4}}+...=-\frac{1}{2\pi }%
\frac{O_{3}}{\mid {z}^{\prime }-{z}\mid ^{4}}+... \\
&&{O_{2.3}^{\prime }(z,\overline{z})}{O_{2.4}^{\prime }(z^{\prime },\overline{z%
}^{\prime })} = -\frac{1}{8}\bigg(\frac{1}{\sqrt{2}\pi }\colon \partial _{%
\overline{z}}\phi _{c+}(\overline{z})(\partial _{z}\phi _{s-}(z))^{2}\colon %
\bigg) \nonumber \\ 
&& \times
\bigg(\frac{1}{\sqrt{2}\pi }\colon \partial _{\overline{z}^{\prime
}}\phi _{c+}(\overline{z}^{\prime })\partial _{z^{\prime }}\phi
_{s-}(z^{\prime })\partial _{\overline{z}^{\prime }}\phi _{s+}(\overline{%
z^{\prime }})\colon \bigg)=  \nonumber \\
&&-\frac{1}{16\pi ^{2}}\bigg(\colon (\partial _{z}\phi _{s-}(z))^{2}\partial
_{z^{\prime }}\phi _{s-}(z^{\prime })\colon -\frac{2\partial _{z}\phi
_{s-}(z)}{(z-z^{\prime })^{2}}\bigg) \nonumber \\ 
&& \times
\bigg(\colon \partial _{\overline{z}%
}\phi _{c+}(\overline{z})\;\;\partial _{\overline{z^{\prime }}}\phi _{c+}(%
\overline{z^{\prime }})\colon -\frac{1}{(\overline{z}-\overline{z}^{\prime
})^{2}}\bigg)\partial _{\overline{z}^{\prime }}\phi _{s+}(\overline{z}%
^{\prime })\simeq  \nonumber \\
&&-\frac{1}{8\pi ^{2}}\frac{\partial _{\overline{z}}\phi _{s+}(\overline{z}%
)\partial _{z}\phi _{s-}(z)}{\mid z-z^{\prime }\mid ^{4}}+...=\frac{1}{2\pi }%
\frac{O_{2s}}{\mid {z}^{\prime }-{z}\mid ^{4}}+... \\
&&{O_{2.7}^{\prime }(z,\overline{z})}{O_{2.6}^{\prime }(z^{\prime },\overline{z%
}^{\prime })} = -\frac{1}{8}\bigg(\frac{1}{\sqrt{2}\pi }\colon \partial
_{z}\phi _{c-}(z)(\partial _{\overline{z}}\phi _{s+}(\overline{z}%
))^{2}\colon \bigg) \nonumber \\ 
&& \times
\bigg(\frac{1}{\sqrt{2}\pi }\colon \partial
_{z^{\prime }}\phi _{c-}(z^{\prime })\partial _{z^{\prime }}\phi
_{s-}(z^{\prime })\partial _{\overline{z}^{\prime }}\phi _{s+}(\overline{%
z^{\prime }})\colon \bigg)=  \nonumber \\
&&-\frac{1}{16\pi ^{2}}\bigg(\colon (\partial _{\overline{z}}\phi _{s+}(%
\overline{z}))^{2}\partial _{\overline{z}^{\prime }}\phi _{s+}(\overline{z}%
^{\prime })\colon -\frac{2\partial _{\overline{z}}\phi _{s+}(\overline{z})}{(%
\overline{z}-\overline{z}^{\prime })^{2}}\bigg) \nonumber \\ 
&& \times
\bigg(\colon \partial _{{z}%
}\phi _{c-}({z})\;\;\partial _{{z^{\prime }}}\phi _{c-}({z^{\prime }})\colon
-\frac{1}{(z-z^{\prime })^{2}}\bigg)\partial _{{z}^{\prime }}\phi _{s-}({z}%
^{\prime })=  \nonumber \\
&&-\frac{1}{8\pi ^{2}}\frac{\partial _{\overline{z}}\phi _{s+}(\overline{z}%
)\partial _{z}\phi _{s-}(z)}{\mid z-z^{\prime }\mid ^{4}}+...=\frac{1}{2\pi }%
\frac{O_{2s}}{\mid {z}^{\prime }-{z}\mid ^{4}}+... \\
&&{O_{2.4}^{\prime }(z,\overline{z})}{O_{2.6}^{\prime }(z^{\prime },\overline{z%
}^{\prime })} = \frac{1}{16}\bigg(\frac{2}{\sqrt{2}\pi }\colon \partial _{%
\overline{z}}\phi _{c+}(\overline{z})\partial _{z}\phi _{s-}(z)\partial _{%
\overline{z}}\phi _{s+}(\overline{z})\colon \bigg) \nonumber \\ 
&& \times
\bigg(\frac{2}{\sqrt{2}%
\pi }\colon \partial _{z^{\prime }}\phi _{c-}(z^{\prime })\partial
_{z^{\prime }}\phi _{s-}(z^{\prime })\partial _{\overline{z}^{\prime }}\phi
_{s+}(\overline{z^{\prime }})\colon \bigg)\simeq  \nonumber \\
&&\frac{1}{8\pi ^{2}}\frac{\partial _{\overline{z}}\phi _{c+}(\overline{z}%
)\partial _{z}\phi _{c-}(z)}{\mid z-z^{\prime }\mid ^{4}}+...=\frac{1}{2\pi }%
\frac{O_{2c}}{\mid {z}^{\prime }-{z}\mid ^{4}}+... \\
&&{O_{2.2}^{\prime }(z,\overline{z})}{O_{2.5}^{\prime }(z^{\prime },\overline{z%
}^{\prime })} = \frac{1}{32\pi ^{2}}\bigg(\colon (\partial _{z}\phi
_{c-}(z))^{2}\partial _{\overline{z}}\phi _{c+}(\overline{z})\colon \bigg)%
\;\;\bigg(\colon \partial _{z^{\prime }}\phi _{c-}(z^{\prime })(\partial _{%
\overline{z}^{\prime }}\phi _{c+}(\overline{z}^{\prime }))^{2}\colon \bigg)=
\nonumber \\
&&\frac{1}{32\pi ^{2}}\bigg(\colon (\partial _{z}\phi _{c-}(z))^{2}\partial
_{z^{\prime }}\phi _{c-}({z}^{\prime })\colon -\frac{2\partial _{z}\phi _{c-}%
}{({z}^{\prime }-{z})^{2}}\bigg)\bigg(\colon \partial _{\overline{z}}\phi
_{c+}(\overline{z})(\partial _{z^{\prime }}\phi _{c+}(\overline{z}^{\prime
}))^{2}\colon -\frac{2\partial _{\overline{z}}\phi _{c+}}{(\overline{z}%
^{\prime }-\overline{z})^{2}}\bigg)\simeq  \nonumber \\
&&\frac{1}{8\pi ^{2}}\frac{\partial _{\overline{z}}\phi _{c+}(\overline{z}%
)\partial _{z}\phi _{c-}(z)}{\mid z-z^{\prime }\mid ^{4}}+...=\frac{1}{2\pi }%
\frac{O_{2c}}{\mid {z}^{\prime }-{z}\mid ^{4}}+...
\end{eqnarray}

\end{document}